\documentclass[reprint,pra,eqsecnum,superscriptaddress]{revtex4-1}
\usepackage[T1]{fontenc}
\usepackage{amsmath}
\usepackage{amssymb}
\allowdisplaybreaks
\usepackage{txfonts}
\let\lambda\lambdaup
\usepackage{array}
\usepackage{graphicx}
\usepackage{xcolor}
\usepackage[colorlinks=true,linkcolor=blue,citecolor=blue,urlcolor=blue]{hyperref}
\allowdisplaybreaks

\newcommand{\Sa}{\mathcal{S}}
\newcommand{\Da}{\mathcal{D}}
\newcommand{\Ma}{\mathcal{M}}
\newcommand{\Za}{\mathcal{Z}}
\newcommand{\ud}{\mathrm{d}}

\begin{document}
\title{Generalised route to effective field theories 
for quantum systems with local constraints}

\author{Attila Szab\'o}
\affiliation{TCM Group, Cavendish Laboratory, University of Cambridge, J. J. Thomson Avenue, Cambridge CB3 0HE, United Kingdom}
\author{Garry Goldstein}
\affiliation{Physics and Astronomy Department, Rutgers University, Piscataway, NJ 08854, USA}
\author{Claudio Castelnovo}
\affiliation{TCM Group, Cavendish Laboratory, University of Cambridge, J. J. Thomson Avenue, Cambridge CB3 0HE, United Kingdom}
\author{Alexei M. Tsvelik}
\affiliation{Condensed Matter Physics and Materials Science Division, Brookhaven National Laboratory, Upton, NY 11973-5000, USA}

\begin{abstract}
Some of the exciting phenomena uncovered in strongly correlated systems 
in recent years -- for instance quantum topological order, deconfined 
quantum criticality, and emergent gauge symmetries -- appear in systems in which 
the Hilbert space is effectively projected at low energies in a way that 
imposes local constraints on the original degrees of freedom. 
Cases in point include spin liquids, valence bond systems, dimer models, and 
vertex models. 
In this work, we use a slave boson description coupled to a large-$S$ path 
integral formulation to devise a generalised route to obtain effective 
field theories for such systems. 
We demonstrate the validity and capability of our approach by studying 
quantum dimer models and by comparing our results with the existing literature. 
Field-theoretic approaches to date are limited to bipartite lattices, they 
depend on a gauge-symmetric understanding of the constraint, and they lack generic 
quantitative predictive power for the coefficients of the terms that appear in 
the Lagrangians of these systems. 
Our method overcomes all these shortcomings and we show how the results up to 
quadratic order compare with the known height description of the square 
lattice quantum dimer model, as well as with the numerical estimate of the 
speed of light of the photon excitations on the diamond lattice. 
Finally, instanton considerations allow us to infer properties of the finite-temperature behaviour in two dimensions.
\end{abstract}
\maketitle
%
%

\section{Introduction}
\label{sec:Introduction}

Low-energy descriptions of strongly correlated many-body systems sometimes 
require the introduction of projected Hilbert spaces where the degrees of 
freedom are subject to local constraints. Notable examples include valence 
bond systems, quantum dimer models, and vertex models. 
The action of the system Hamiltonian within the restricted Hilbert space 
often gives rise to exotic and unexpected behaviour, from emergent gauge 
symmetries and deconfined quantum criticality to quantum topological 
order, which have been the subject of much attention in recent years. 

Field-theoretic descriptions of these systems have proven to be a powerful 
tool to study their properties, in particular to understand the nature of 
their correlations and critical points. However, conventional routes 
to construct such field theories often do not apply, and instead 
ad hoc methods have been devised throughout the years. 
Such methods largely hinge on a physical understanding 
of the constraint and how to best represent it in a (free) field 
theory language. 
While on the one hand such approaches have provided great insight into the 
relevant systems, they cannot easily be generalised. A systematic way to 
arrive at a field-theoretic description of quantum systems with local 
constraints is currently lacking. 

In this paper, we propose a generalised route to obtain field-theoretic 
actions from microscopic Hamiltonians based on a slave boson 
representation of the relevant degrees of freedom and their constraints, 
combined with a large-$S$ path integral formulation. 
We demonstrate the validity and capability of our approach by deploying 
it to study quantum dimer models (QDMs), recovering known 
and obtaining new results on bipartite lattices, and showing that it can be 
used straightforwardly on hitherto inaccessible non-bipartite lattices. 
Our approach also paves the way to semiclassical simulations of these systems, 
as discussed in Ref.~\onlinecite{Szabo2019} by some of the authors.

QDMs were 
introduced to describe a magnetically disordered 
(resonating valence bond) phase in high-temperature 
superconducting materials~\cite{Rokshar1988}. 
They can also arise in Bose--Mott insulators, electronic Mott insulators
at fractional fillings~\cite{Lee2003}, and in mixed valence systems
on frustrated lattices~\cite{Fulde2002}. 
For a review of these models, we refer the reader to 
Ref.~\onlinecite{Moessner2011}. 

When QDMs are defined on bipartite lattices, they are amenable to a height 
mapping description in two dimensions (2D), which generalises to quantum electrodynamics (QED) 
in 3D~\cite{Huse2003,Moessner2003,Moessner2011,Fradkin_2013}. 
The height mapping is built upon a gauge-symmetric understanding 
of the constraint~\cite{Henley2010} and has great predictive power for 
bipartite quantum dimer models, enabling one to answer detailed 
questions pertaining to their long-wavelength properties. 
All its good features notwithstanding, the height mapping has one essential 
drawback: a quantitative derivation of its action from the 
microscopic lattice Hamiltonian is currently not available.
The state-of-the-art derivation of the height mapping has been so far  
phenomenological~\cite{Henley1997,Moessner2011}, 
and the coefficients (including their signs) are {\it a posteriori} determined by theoretical 
considerations (e.g., using the knowledge of the exact ground state at fine-tuned critical points) 
combined with comparisons to numerical results~\cite{Henley1997,Huse2003,%
Moessner2003,Fradkin2004,Moessner2011,Fradkin_2013}. 
Furthermore, the approach cannot be applied to nonbipartite lattices, which remain 
comparatively unexplored from a field-theoretic perspective. 

Using our generalised route, we show that one can systematically derive the 
height Lagrangian for bipartite QDMs in 2D and 3D, from the corresponding 
microscopic Hamiltonians. We compare our results against the existing 
literature, where we find good agreement considering the large-$S$ and 
quadratic approximations that we employ. 
For example, we derive the stiffness of the 2D square lattice QDM 
at a well-known critical point called the Rokhsar--Kivelson (RK) point~\cite{Rokshar1988}, 
where the ground-state wave function of the system is known exactly:
Our result, $1/4$, is comparable to the exact value, $\pi/18$~\cite{Moessner2011,Tang2011}. 
We use instanton considerations to discuss the fate of 2D phases at finite temperature.
We also obtain the speed of light to quadratic order in 
the 3D QED long-wavelength theory of dimers on the cubic lattice, 
$c=\sqrt{2J\left(J-V\right)}\,S/3$, 
and on the diamond lattice, $c=\sqrt{J(J-V)/6}\,\,S^2$. 
The latter is known numerically from quantum Monte Carlo simulations 
to be $c(S=1) \simeq \sqrt{0.8 J(J-V)}$~\cite{Sikora2009,Sikora2011}. 
We further show that our approach applies straightforwardly to the 
non-bipartite case of the QDM on the triangular lattice, where we observe a 
curious analytical similarity with the formalism for the QDM on the 3D cubic 
lattice -- a similarity that we plan to explore further in future work. 

The paper is organised as follows. 
In Sec.~\ref{sec:Basic-formulation}, we introduce our approach by studying 
in detail the introductory and well-known case of the QDM on the square 
lattice. 
In Sec.~\ref{sec:Basic-formulation_cubic}, we study the cubic 
lattice and obtain the dispersion of its photon excitations in the gapless 
phase. We then briefly consider the case of the nonbipartite triangular 
lattice in Sec.~\ref{sec:triangular-lattice}, which is shown to be curiously 
similar in its analytical form to the cubic lattice. 
Finally, in Secs.~\ref{sec:Hexagonal-lattice} and~\ref{sec:Diamond-lattice}, we consider for completeness 
the QDMs on the honeycomb and diamond lattices, respectively. 
A comparison with the conventional height mapping and gauge theoretic
formulations is presented in Sec.~\ref{sec: other approaches}, 
and we conclude in Sec.~\ref{sec:Conclusions}. 
%
%

\section{Square lattice}
\label{sec:Basic-formulation}

\begin{figure}
\includegraphics{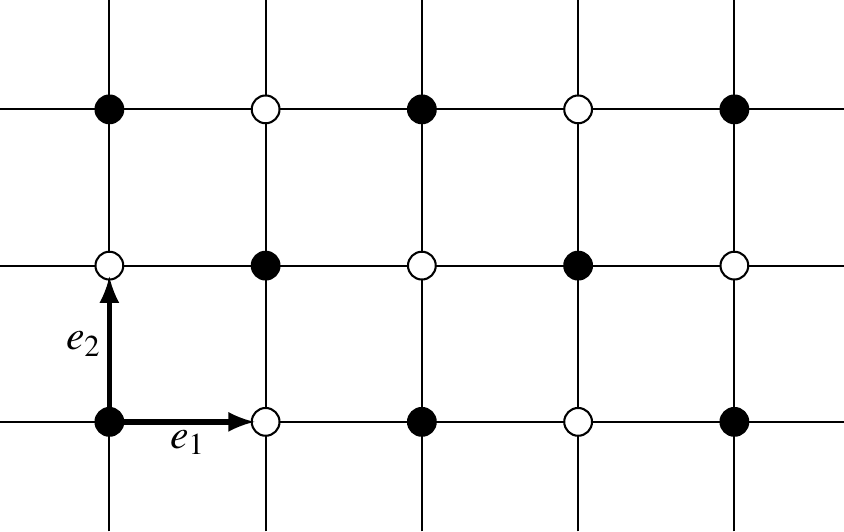}
\caption{Square lattice showing the choice of basis vectors $e_{1,2}$.}
\label{fig:square lattice}
\end{figure}

\subsection{Bosonic representation and large $S$}
\label{subsec:RK-Hamiltonian}

We consider in the first instance the well-known case of the RK Hamiltonian 
for the square lattice QDM, which allows us to introduce 
our approach in the simplest setting. Generalisations to the honeycomb, 
triangular, cubic and diamond lattices will be discussed in later sections. 
%
%

The quantum dimer model can be mapped exactly onto a slave boson model
by considering a secondary Hilbert space where we assign a bosonic
mode $b_{r,\eta}$ to each link ($r,r+e_\eta$) of the square lattice ($e_\eta=\hat{x},\hat{y}$)~\cite{Goldstein_2017} (see Fig.~\ref{fig:square lattice}). 
We associate the number of dimers on a link with the occupation number of the bosons on that link, thus embedding the dimer Hilbert space in the larger Hilbert space
of the bosons. 

The constraint that each site of the lattice has one and exactly one dimer attached to it can be expressed in the boson language as
\begin{equation}
\Pi_{r}\equiv\sum_{l\in v_{r}}b_{l}^{\dagger}b_{l}^{\phantom\dagger}-1=0. 
\label{eq:Constraint}
\end{equation}
Here, for convenience of notation, $\ell\in v_{r}$ labels the four
links $r',\eta$ that are attached to the vertex $r$. We note that the
constraint in Eq.~\eqref{eq:Constraint} implies that the bosons are
hard-core: $n_{r,\eta} \equiv b^{\dagger}_{r,\eta} b_{r,\eta} = 0,1$. 

Any dimer Hamiltonian has a bosonic representation; 
in particular, the RK Hamiltonian can be written as 
\begin{align}
H_{D}= & 
\sum_{r}\left\{ 
  -J\,b_{r,1}^{\dagger}b_{r+e_2,1}^{\dagger}
	    b_{r,2}^{\;}b_{r+e_1,2}^{\;}
	+\:(1 \leftrightarrow 2) 
\right.
\nonumber \\*
 & \left.
\qquad 
   +V\,b_{r,1}^{\dagger}b_{r,1}^{\;}
	     b_{r+e_2,1}^{\dagger}b_{r+e_2,1}^{\;}
	 +\:(1 \leftrightarrow 2) 
 \right\} 
. 
\label{eq:RK_Hamiltonian}
\end{align}
To generalise this construction to a large-$S$ formulation, we keep the
same Hilbert space as before and replace Eq.~\eqref{eq:Constraint} 
with the constraint
\begin{equation}
\Pi_{r}\equiv\sum_{l\in v_{r}}b_{l}^{\dagger}b_{l}^{\;}-S=0 
\label{eq:Constraint_S}
\end{equation}
without changing the Hamiltonian~\eqref{eq:RK_Hamiltonian}.
%
%

\subsection{Path integral formulation and Gaussian approximation}
\label{sec:Path-integral-formulation}

In what follows, it is convenient to use the radial gauge
for the bosonic fields in the path integral formulation of the model:
\begin{eqnarray}
b_{r,\eta}
&=&
\sqrt{\rho_\eta(r+e_\eta/2)}
  \exp\left[ i \Phi_\eta(r+e_\eta/2) \right]
\label{eq:radial_representation}
\\*
&\equiv&
\sqrt{\frac{S}{z}+\delta\rho_\eta(r+e_\eta/2)}
  \exp\left[ i \Phi_\eta(r+e_\eta/2) \right] 
, 
\nonumber 
\end{eqnarray}
where $z$ is the lattice coordination number; for the square lattice, 
$z =4$. For later convenience, we think of $\rho_\eta(r+e_\eta/2)$ and 
$\Phi_\eta(r+e_\eta/2)$ as functions defined on bond midpoints. 
The RK Hamiltonian is then given by
\begin{widetext}
\begin{align}
H_{D}= & \sum_{r}
\Big\{ 
  -2J \, \sqrt{\rho_1(r+e_1/2) \rho_2(r+e_1+e_2/2) 
	             \rho_1(r+e_2+e_1/2) \rho_2(r+e_2/2)}
\nonumber \\*
& \qquad
\times
\cos\left[
  \Phi_1(r+e_1/2)-\Phi_2(r+e_1+e_2/2)+\Phi_1(r+e_2+e_1/2)-\Phi_2(r+e_2/2)
\right]
\nonumber \\*
& \qquad
+ V \, \rho_1(r+e_1/2) \rho_1(r+e_2+e_1/2)
+ V \, \rho_2(r+e_2/2) \rho_2(r+e_1+e_2/2)
\Big\} 
,
\label{eq:RK_Hamiltonian_radial}
\end{align}
\end{widetext}
and the constraint in Eq.~\eqref{eq:Constraint_S} can be written as
\begin{equation}
\sum_\eta 
\left[ 
  \delta\rho_\eta(r+e_\eta/2)+\delta\rho_\eta(r-e_\eta/2) 
\right] 
= 0
\, . 
\label{eq:Cobstraint_solution}
\end{equation}
We now introduce the Fourier decomposition
\begin{align}
\delta\rho_\eta(r+e_\eta/2) & 
=\frac{1}{\sqrt{N}}\sum_{k}
  \delta\rho_{\eta}(k) \exp\left[ -i k (r + e_\eta/2) \right] ,
\label{eq:Fourier_transform} \\ 
\Phi_\eta(r+e_\eta/2) & 
=\frac{1}{\sqrt{N}}\sum_{k} 
\Phi_{\eta}(k) \exp\left[ -i k (r + e_\eta/2) \right] ,
\end{align}
where $k (r + e_\eta/2) \equiv \vec{k}\cdot(\vec{r}+\vec{e}_{\eta}/2)$ for brevity and
$N$ is the number of lattice sites. In these terms, the constraint can be 
written as 
\begin{equation}
\sum_\eta \cos(k e_\eta/2) \, \delta\rho_\eta(k) = 0 .
\label{eq:k_space_constraint}
\end{equation}
It will be useful in the following to introduce the shorthand notation 
$c_\eta = \cos(k e_\eta/2)$ and $s_\eta = \sin(k e_\eta/2)$. 

The constraint clearly imposes a relation between the two field variables 
$\delta\rho_1(k)$ and $\delta\rho_2(k)$. The same conclusion can be readily 
drawn about the fields $\Phi_\mu$ once we notice that the Hamiltonian depends 
only on the specific combination of them that appears in the argument of the 
cosine term in Eq.~\eqref{eq:RK_Hamiltonian_radial}: 
\begin{eqnarray}
\tilde\phi(r) &\equiv& 
\Phi_1(r+e_1/2)-\Phi_2(r+e_1+e_2/2)
\nonumber \\* 
&&+ \Phi_1(r+e_2+e_1/2)-\Phi_2(r+e_2/2) ,
\end{eqnarray}
whose Fourier transform is
\begin{eqnarray}
\tilde\phi(k) &=& 
e^{-ik(e_1+e_2)/2} \, 2 \left[ 
  c_2 \Phi_1(k) - c_1 \Phi_2(k) 
\right] .
\end{eqnarray}
Note that the cosine function depends only on powers of $\phi(r)^2$ and 
therefore phase factors in $\tilde\phi(k)$ are immaterial, and we define 
for convenience 
\begin{eqnarray}
\phi(k) &\equiv& e^{ik(e_1+e_2)/2} \, \tilde\phi(k) 
\nonumber \\*
&=& 
2 \left[ 
  c_2 \Phi_1(k) - c_1 \Phi_2(k) 
\right]
\equiv 
\Za_\eta \Phi_\eta,
\label{eq: phi(k) square lattice}
\end{eqnarray}
where $\Za_\eta = (2 c_2, -2 c_1)$. In real space, this amounts to introducing a $\phi(r)$ living on the centres of the plaquettes rather than a corner.

Notice that the constraints on $\delta\rho_\eta$ and on $\Phi_\eta$ 
are in fact two sides of the same coin -- indeed, conjugate variables come in pairs, so their numbers have to be the same. 
In our case, one can easily verify that imposing one of them implies the other. 
This is a consequence of how the RK Hamiltonian is designed: the plaquette-flipping term inherently preserves 
the number of dimers at each vertex; and vice versa, if one imposes the 
hard core dimer constraint, then any kinetic contribution in the Hamiltonian 
is projected onto a combination of loop updates, of which the 
plaquette-flipping term is an example. 

To make further progress in the path integral formulation, we shall 
expand the action to quadratic order in $\phi(r,\tau) \simeq 2 n \pi$, $n \in \mathbb{Z}$, 
and $\delta \rho_{r,\mu} \ll 1$. 
Firstly, it is convenient to rewrite $\cos(\phi) = 1 - [1-\cos(\phi)]$ and 
notice that the term in square brackets contains only quadratic and higher-order contributions. 
Therefore, the square root in the second term of
\begin{equation}
\sqrt{\rho\rho\rho\rho}\cos(\phi)
= 
\sqrt{\rho\rho\rho\rho}
-
\sqrt{\rho\rho\rho\rho}
\left[1 - \cos(\phi)\right] 
,
\label{eq: rho cos expansion square}
\end{equation}
needs to be expanded only to leading order in $S$: 
$\sqrt{\rho\rho\rho\rho} \simeq S^2 / 16$. 
Upon expanding the first term, one obtains both 
linear and quadratic terms in $\delta \rho_\eta(r+e_\eta/2)$. 
However, one can readily 
convince oneself that the linear terms vanish upon summing over $r$ because 
of the dimer constraint~\eqref{eq:Cobstraint_solution}. 
The same is true for the linear contributions due to the terms multiplying $V$ in Eq.~\eqref{eq:RK_Hamiltonian_radial}, leading to the following contributions to quadratic order: 
\begin{widetext}
\begin{eqnarray}
\sqrt{\rho\rho\rho\rho} 
&\simeq& 
\frac{S^2}{16} 
+ 
\frac{1}{4} \Big[ 
  \delta \rho_1(r+e_1/2) \delta \rho_2(r+e_1+e_2/2) 
  + \delta \rho_1(r+e_1/2) \delta \rho_1(r+e_2+e_1/2) 
\label{eq: rho expansion J}
\\*
&& \qquad \quad
  + \delta \rho_1(r+e_1/2) \delta \rho_2(r+e_2/2) 
  + \delta \rho_2(r+e_1+e_2/2) \delta \rho_1(r+e_2+e_1/2) 
\nonumber \\*
&& \qquad \quad
  + \delta \rho_2(r+e_1+e_2/2) \delta \rho_2(r+e_2/2)
  + \delta \rho_1(r+e_2+e_1/2) \delta \rho_2(r+e_2/2)
\Big] 
\nonumber \\
&-& 
\frac{1}{8} \Big[ 
  \delta \rho_1(r+e_1/2)^2 
	+\delta \rho_1(r+e_2+e_1/2)^2
	+\delta \rho_2(r+e_1+e_2/2)^2 
	+\delta \rho_2(r+e_2/2)^2
\Big] 
\nonumber \\
\rho\rho + \rho\rho
&=& 
\frac{S^2}{8} 
+ \delta \rho_1(r+e_1/2) \delta \rho_1(r+e_2+e_1/2) 
+ \delta \rho_2(r+e_2/2) \delta \rho_2(r+e_1+e_2/2) 
\, . 
\label{eq: rho expansion V}
\end{eqnarray}
\end{widetext}
Writing the sum of these terms in Fourier space, we get 
(for the quadratic contributions only): 
\begin{eqnarray}
&-& J c_1 c_2 \, \left[ 
  \delta\rho_1(k) \delta\rho_2(-k) 
  + 
	\delta\rho_2(k) \delta\rho_1(-k) 
\right] 
\nonumber
\\* 
&+& \left[
  (2V-J) c_2^2 + (J-V)
\right] \delta\rho_1(k) \delta\rho_1(-k) 
\nonumber \\* 
&+& \left[
  (2V-J) c_1^2 + (J-V)
\right] \delta\rho_2(k) \delta\rho_2(-k) 
\, . 
\label{eq: Dk rho square lattice}
\end{eqnarray}
The dynamics of the model is generated by the standard bosonic Berry phase $\sum_n \overline{b_n}\partial_\tau b_n$. In the radial representation \eqref{eq:radial_representation}, this gives rise to the term $\sum_n i\,\delta\rho_n\,\partial_\tau\Phi_n$, as well as total derivative terms that do not contribute to the action. Altogether, we obtain
\begin{eqnarray}
{\Sa} &=& 
\int \!\!\ud\tau \sum_{k,\mu} \: 
  i \, \delta\rho_\mu(k,\tau)\;\partial_{\tau}\Phi_\mu(-k,\tau)
\nonumber \\*
&+& \int \!\!\ud\tau \sum_r \: 
	\frac{JS^{2}}{8} \left[ 1 - \cos(\phi(r,\tau)) \right] 
\nonumber \\*
&+& \int \!\!\ud\tau 
\sum_{k,\mu,\nu} \frac{{\Da}_{\mu\nu}(k)}2 \: 
\delta\rho_\mu\left(k,\tau\right)
\delta\rho_\nu\left(-k,\tau\right) ,
\label{eq: action drho,Phi}
\end{eqnarray}
where 
\begin{eqnarray}
{\Da}_{\mu\nu} &=& 2
\left(
\begin{array}{cc}
J-V + (2V-J) c_2^2 & - J c_1 c_2 
\\ 
- J c_1 c_2 & J-V + (2V-J) c_1^2
\end{array}
\right).
\nonumber 
\end{eqnarray}

To proceed further, we can  either resolve the constraint explicitly, or keep it implicit. 
The former allows to relate directly with the customary height field 
representation for the QDM on the square lattice; the latter is more 
concise and will be useful to reduce the algebra and obtain analytic 
results for three dimensional models in 
Secs.~\ref{sec:Basic-formulation_cubic} and~\ref{sec:Diamond-lattice}. 
For this reason, we present them both in the following sections. 
%
%

\subsection{Implicit constraint}
\label{sec:Fluctuations1}

Let us consider a given cosine minimum at first, and assume 
$\phi(r,\tau) \ll 1$. (We shall discuss the effect of instantons later in 
Sec.~\ref{sec:Instantons}.) 
The middle term in Eq.~\eqref{eq: action drho,Phi} then reduces to 
\begin{equation}
\sum_r \frac{J S^2}{16} \phi^2(r,\tau) 
= \sum_k \frac{J S^2}{16} \Za_\mu \Za_\nu \Phi_\mu(k,\tau) \Phi_\nu(-k,\tau) 
\, . 
\end{equation}
After integrating the first term in  Eq.~\eqref{eq: action drho,Phi} by parts in $\tau$, one can integrate the fields $\Phi_\mu$ out and obtain an action only in terms of the 
fields $\delta\rho_\mu$: 
\begin{align}
{\Sa} = \frac12 \int \!\!\ud\tau \sum_{k} &
	\Big[ 
	\left( \mathcal{M}^{-1} \right)_{\mu\nu}
	  \partial_{\tau}\delta\rho_\mu(k,\tau)
	  \partial_{\tau}\delta\rho_\nu(-k,\tau)
\nonumber \\*
&+  {\Da}_{\mu\nu}(k) \: 
\delta\rho_\mu\left(k,\tau\right)
\delta\rho_\nu\left(-k,\tau\right) \Big],
\label{eq: action drho}
\end{align}
where we define $\Ma \equiv J S^2 \, \Za \Za^T/8$ for convenience.
One can now readily obtain the dispersion by computing the eigenvalues of 
the matrix $\mathcal{M}\mathcal{D}$. 

This approach gives us two modes while we know that the physical system is 
constrained to only one real scalar field. Fortunately, Eq.~\eqref{eq:k_space_constraint} states that $\sum_\mu \Za_\mu \delta\rho_\mu(k) = 0$. Therefore, one of the two eigenvalues of $\mathcal{MD}$ corresponds to an unphysical mode and vanishes, whereas the other (finite) eigenvalue corresponds  to the physical dispersion of the system, 
\begin{equation}
\omega^2 = J S^2 \Big[ 
	(2V-J)(c_1^4+c_2^4)
  + (J-V)(c_1^2+c_2^2)  
	+ 2 J c_1^2 c_2^2
\Big] . 
\label{eq: dispersion QDM square}
\end{equation}

\begin{figure}
\includegraphics{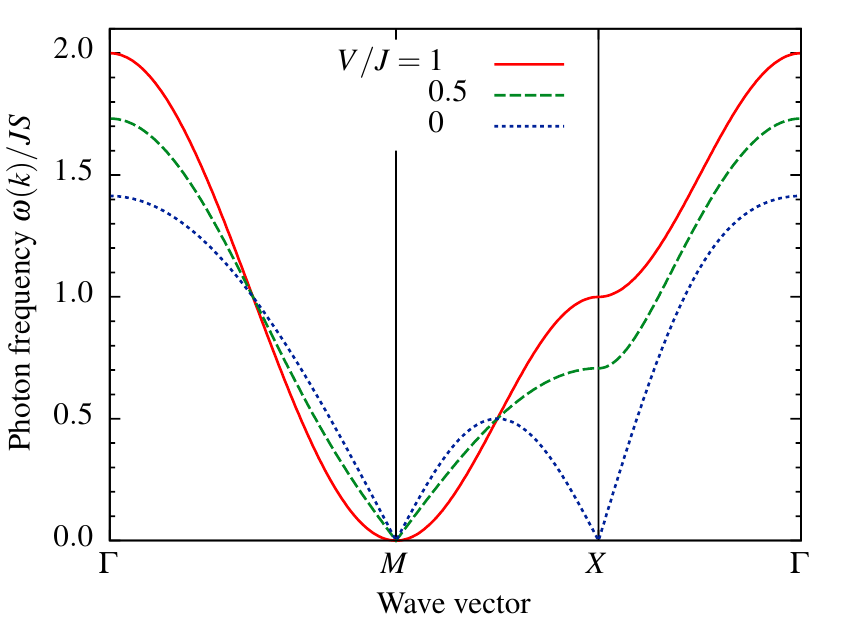}
\caption{Photon dispersion relation of the large-$S$ QDM on the square lattice. The spectrum is gapless at the $M=(\pi,\pi)$ point for all values of $V/J$; near this point, the dispersion is quadratic at the RK point, and linear away from it. Another minimum forms at the $X=(\pi,0)$ points for lower $V$: this drives an ordering transition at $V=0$. }
\label{fig:square dispersion}
\end{figure}

This dispersion is plotted for three values of $V/J$ in 
Fig.~\ref{fig:square dispersion}.
It is interesting to note that the dispersion vanishes at $(\pi,\pi)$ and 
symmetry related points for all values of $J$ and $V$. 
An instability develops for $V>J$ when the dispersion becomes negative 
near the $(\pi,\pi)$ point (not shown). 
As we lower the value of $V$, secondary minima appear at $(\pi,0)$ and 
related points in the Brillouin zone, and they drive the system through 
an instability for $V < 0$ which leads to (plaquette) dimer ordering at these wave vectors. 
%
%

\subsection{Explicit constraint}
\label{sec:Fluctuations2}

The implicit approach in Sec.~\ref{sec:Fluctuations1} allows us to arrive at 
the dispersion of the system with minimal algebra, but it does not produce an 
action in terms of the physical degree of freedom only. 
In order to achieve this, we need to resolve the constraint explicitly. 
A convenient way to do so is to look for a conjugate field $h(k)$ such that 
the Berry phase in 
the path integral can be written as $i h(k) \partial_\tau \phi(-k)$:
\begin{eqnarray}
h(k) \partial_\tau \phi(-k) 
= 
\sum_\eta \delta\rho_\eta(k) \partial_\tau \Phi_\eta(-k) .
\end{eqnarray}
Substituting the expression for $\phi(k)$, 
Eq.~\eqref{eq: phi(k) square lattice}, into the equation above, we obtain 
\begin{align}
\delta\rho_1(k) &= 2 c_2 \, h(k) ,
&
\delta\rho_2(k) &= -2 c_1 \, h(k) .
\label{eq:square lattice resolved constraint}
\end{align}
One can then straightforwardly verify that introducing the field 
$h(k)$ automatically resolves the constraint:
\begin{eqnarray}
\sum_\eta c_\eta \delta\rho_\eta = 
2(c_1 c_2 - c_2 c_1) h(k) = 0 .
\end{eqnarray}
Once again, this result should not come as a surprise. It is a reflection, 
at a field-theoretic level, of the fact that the plaquette terms in the Hamiltonian 
respect the dimer constraint. Therefore, if the field theory is built from 
plaquette kinetic terms only, then the constraint is implied. 

We are thus in the position to write the full large-$S$ action for the system, 
including both the Berry phase and Hamiltonian contributions, in terms of 
the fields $h(k)$ and $\phi(k)$ only. 
Adding more complicated ring exchange type terms to the RK 
Hamiltonian does not invalidate this conclusion, as the phase in each ring 
exchange term may be written as a sum of phases over single plaquettes. 
Substituting the expressions of $\delta\rho_\eta(k)$ in terms of $h(k)$, 
and ignoring trivial constants, we obtain the action: 
\begin{eqnarray}
{\Sa} &=& \!\int \!\!\ud\tau \sum_r 
\left\{
  i h(r,\tau)\partial_{\tau}\phi(r,\tau)
	+ \frac{JS^{2}}{8} \big[ 1 - \cos(\phi(r,\tau)) \big] 
\right\} 
\nonumber \\*
&+& \int \!\!\ud\tau 
\sum_{k} \frac{{\Da}_0(k)}2 h(k,\tau)\,h(-k,\tau) 
\label{eq: action h,phi}
\end{eqnarray}
where 
\begin{equation}
{\Da}_0(k) = 8 \Big[
  (2V-J)(c_1^4 + c_2^4) 
	+ (J-V)(c_1^2 + c_2^2)
	+ 2 J c_1^2 c_2^2
\Big] . 
\label{eq:Action_final}
\end{equation}
%
%

\subsubsection{Action without instantons}
\label{subsec:Action-without-instantons}

Ignoring for the time being the contribution to the action due to instantons 
between different minima of the cosine term, we can expand about one given 
minimum and integrate over $\phi$ to arrive at 
\begin{align}
\Sa = \frac12 \int \!\!\ud\tau \: \sum_{k}{}&
\Big[ \frac{8}{JS^{2}}\partial_{\tau}h(k,\tau)\partial_{\tau}h(-k,\tau)
\nonumber \\*
&  +{\Da}_0(k)h(k,\tau)h(-k,\tau)\Big].
\label{eq:General_form_h_square}
\end{align}
One can easily see that the dispersion is indeed the same as in 
Eq.~\eqref{eq: dispersion QDM square}. 

Expanding ${\Da}_0$ about its minimum at $\left(\pi,\pi\right)$, 
\begin{eqnarray}
{\Da}_0[(\pi,\pi)+(k_x,k_y)] &\simeq& 
  2(J-V)(k_x^2 + k_y^2)
\nonumber \\* 
&+& 
	\left[ \frac{7 V}{6} - \frac{2J}{3} \right] (k_x^4 + k_y^4)
	+ J k_x^2 k_y^2
\, , 
\nonumber 
\end{eqnarray}
we obtain the action
\begin{align}
{\Sa} =\frac12& \int \!\! \ud\tau\, \ud^2r\,
 \bigg\{\frac{8}{JS^{2}}(\partial_{\tau}h)^{2}+2(J-V)(\nabla h)^{2}
\label{eq:Action_2d_final}
\\*
& +\frac{7V-4J}{6}
    h\left(\partial_{x}^{4}+\partial_{y}^{4}\right)h
	+ J h\left(\partial_{x}^{2}\partial_{y}^{2}\right)h 
\bigg\}+\dots
\nonumber 
\end{align}
At the RK point, the $(\nabla h)^{2}$ term
vanishes and the terms with quartic derivatives add up to 
\begin{equation}
\frac{J}{2}(\nabla^{2}h)^{2} 
\, , 
\end{equation}
yielding the spectrum $\omega = k^2/2m$ with $m = 2/JS$. 
We note that the known value at the RK point for 
$S=1$ (and $J=V=1$) in this normalisation is $m = 9/\pi$ 
(corresponding to $K=\pi/18$ in 
Refs.~\onlinecite{Moessner2011,Tang2011}), 
which can be obtained exactly from the ground-state 
wave function of the QDM, available only at the RK point). 
This shows that, expanding to quadratic order, our estimate is within 
40\% accuracy. 
We note that such a discrepancy at quadratic order in a large-$S$ 
expansion is consistent for instance with similar results obtained in 
Ref.~\onlinecite{Kwasigroch2017} for quantum spin ice. 
Our results can be improved by going to higher orders, and -- more 
importantly with respect to earlier work on field theories for quantum dimer 
models -- their validity is not limited to the fine tuned RK point. 
%
%

\subsubsection{Instantons}
\label{sec:Instantons}

We will now incorporate the instanton effects which, as we shall demonstrate, 
always generate a mass for the photons for $V<J$, as it generally happens in 
compact electrodynamics. To this end, we are 
going to integrate out the field $\phi\left(r,\tau\right)$ 
taking into account the fact that the action is periodic in it. 

Firstly, we proceed by the standard Villain approach and replace 
\begin{eqnarray}
1-\cos\phi
\;\; 
\rightarrow \;\; 
\frac{1}{2}\Big[\phi-2\pi\sum_{j}q_{j}\theta(\tau-\tau_{j})\Big]^{2}\!\!, \label{eq:Villain}
\end{eqnarray}
where the $q_{j}=q(r_j,\tau_j)$ are integers representing  instanton events, and $\theta(\tau)$ is the 
Heaviside step function. 
By integrating over $\phi$ and $h$, we obtain the following action: 
\begin{eqnarray}
&& {\Sa} = 
\frac{(2\pi)^{2}}{2}\sum_{j,k}
q(r_{j},\tau_j)G_{qq}(r_{j}-r_{k};\tau_j-\tau_k)q(r_{k},\tau_k)
\nonumber \\
&& G_{qq}(k,\omega) = 
\left[\omega^{2}/M + (\rho_{2}k^{2}+\rho_{4}k^{4})\right]^{-1}
\, , 
\label{eq: CG action Villain instanton}
\end{eqnarray}
where $M=JS^{2}/8$ and we introduced the symbolic terms $\rho_{2}k^{2}$ 
and $\rho_{4}k^{4}$ to represent the quadratic and quartic derivative terms 
in the action: $\Da_0(k) \simeq \rho_2k^2+\rho_4k^4+\dots$
The resulting partition function is that of a Coulomb gas of charges 
$q = \pm 1, \pm 2, \dots$, where the fugacity of charge $q$ 
is given by  $I = \exp(-q^2 {\Sa}_0)$ and 
${\Sa}_0$ is the contribution to the action from a single instanton with 
$q=1$: 
\begin{equation}
\Sa_0 = \frac1{4\pi} \int\!\frac{\ud\omega\, \ud^{2}k}
{\omega^{2}/M+(\rho_{2}k^{2}+\rho_{4}k^{4})}  \approx 
\frac{\pi S}{8}\sqrt{\frac{J}{2\rho_4}}\,\ln(\rho_{4}/\rho_{2}) ,
\label{eq:single instanton action}
\end{equation}
and hence
\begin{equation}
I = (\rho_{2}/\rho_{4})^{q^2 \pi S/8},
\label{eq:I}
\end{equation}
where we performed the calculations with the RK form of the quartic term 
and substituted $\rho_4=J/2$ for simplicity. 

Since $I$ is a rapidly decaying function of the instanton charge, we can restrict 
our consideration to the gas of charges $q=\pm 1$. 
Following Polyakov~\cite{Polyakov1975},
we approximate the partition function of the Coulomb 
gas~\eqref{eq: CG action Villain instanton} as the one of the sine-Gordon 
model with  action~\eqref{eq:Action_2d_final} augmented by the term 
\begin{equation}
\delta S = -2\mu I \! \int \ud\tau \sum_r \cos(2\pi h),
\label{eq:z}
\end{equation}
where $\mu\ud\tau$ is the preexponential part of the instanton measure 
(see Appendix~\ref{app: instanton}); at the RK point, 
$2\mu = JS^{3/2}\sqrt{\pi/2}$. 
The presence of this term makes the excitations massive:
\begin{equation}
\omega^2 = c^2k^2+ m^2, 
\qquad 
m^2 = 8\pi^2 M\mu I.
\end{equation}
As we see from~\eqref{eq:I}, this mass vanishes at the RK point. 

At finite temperatures, instantons interact logarithmically:
\begin{equation}
E = -\frac{2\pi T^2}{\rho_2} \sum_{j<k} q_j q_k \ln\left(\frac{|r_j-r_k|}{r_0}\right), 
\end{equation}
where $r_0^2 = \rho_4/\rho_2$. 
The corresponding contribution to the free energy density 
\begin{equation}
\delta F \propto q^2 I_q \int \frac{\ud^2r }{r^{d_q}}
\, ,
\qquad 
d_q = 2\pi q^2T/\rho_2,
\label{F}
\end{equation}
diverges at small $T$. 
The contribution of the static fluctuations of $h$ is 
encoded in the free energy functional 
\begin{eqnarray}
F =
\int \!\!\ud^2 r \: 
\left[ 
\frac{\rho_2}{2}(\nabla h)^2 
+ \frac{\rho_4}{2}(\nabla^2 h)^2 
- 2\mu I\cos(2\pi h) 
\right] .
\nonumber \\
\label{sine-G}
\end{eqnarray}
The scaling dimension $d_1$ of the cosine term is given by~\eqref{F}. 
The critical temperature above which the cosine term is irrelevant is 
determined by the condition $d_1=2$: 
\begin{equation}
T_c = \rho_2/\pi .
\end{equation}
Above $T_c$, we have a critical phase; below it, the correlation length 
of the $h$ field is finite:
\begin{equation}
	\xi\sim r_0 (\mu I/ T_c)^{-1/(2-d_1)}
\end{equation}
This  corresponds to melting the valence bond crystal. 
As it is typical for 2D crystals, it melts via a 
Berezinskii--Kosterlitz--Thouless transition. 

We finally take a moment to comment on the difference between our result and
the one obtained in Ref.~\onlinecite{Papanikolaou2007}, which in fact addresses
a somewhat different problem. In Ref.~\onlinecite{Papanikolaou2007} the
authors considered equal time correlations at the RK point. The remarkable
property of this point is that the ground-state wave function can be
represented as a path integral with a  Gaussian action. It was argued that
the periodicity of the height field generates the irrelevant perturbation
$\cos(2\pi h)$. Here, we obtained a formally equivalent perturbation~\eqref{eq:z}
also away from the RK point; however, the prefactor of the cosine term in our
case vanishes precisely at the RK point. 
%
%

\subsection{Large-$S$ phase diagram}
\label{sec: square classical phase diagram}

\begin{figure}
	\includegraphics{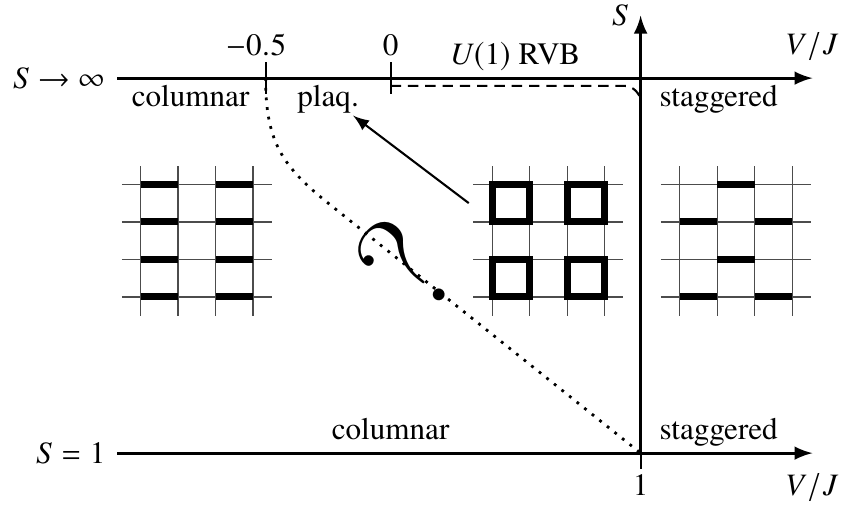}
	\caption{Phase diagram of the square lattice quantum dimer model at $S=\infty$ (this work) and $S=1$~\cite{Sachdev1989,Leung1996,Zeng1997,Syliuasen2006,Ralko2008,Banerjee2014,Oakes2018,HerzogArbeitman2019}. }
	\label{fig:square phase diagram}
\end{figure}

As $S\to\infty$, zero point fluctuations of any soft modes are negligible, and $\rho$ and $\Phi$ can be treated as commuting, classical variables. 
The ground state energy of the system in this limit is therefore given by 
classical minimisation of the Hamiltonian~\eqref{eq:RK_Hamiltonian_radial}. 
Since $\sqrt{\rho\rho\rho\rho}\ge 0$ always, the $\Phi$ in such an optimal 
state satisfy $\cos(\Phi_1-\Phi_2+\Phi_3-\Phi_4)=1$ which is achieved, 
for instance, by setting $\Phi\equiv0$.

Finding the optimal values of $\rho$ in full generality is more difficult. However, one can always compare the ground state energies of phases suggested in the literature, or develop a variational ansatz that captures several such phases. 
In the case of the square lattice, we considered states in which $\rho$ is constant within each set of bonds populated in the four columnar ordered states. Such an ansatz can capture columnar and plaquette ordered states as well as the RVB liquid phase. 

Comparing the ground state energies of these phases yields the $S\to\infty$ phase diagram shown in Fig.~\ref{fig:square phase diagram}. As expected, the ground state is staggered for $V>J$ and columnar at $V\to-\infty$. At intermediate $V/J$, we see a plaquette ordered phase as well as an extended $U(1)$ RVB liquid, with phase boundaries corresponding to the instabilities shown in Fig.~\ref{fig:square dispersion}. The latter is unstable at finite $S$ due to instanton effects, as discussed above. 
The fate of the plaquette phase is, however, less clear: since it is ordered, instantons are unlikely to substantially affect its stability, and so the evolution of the columnar--plaquette phase boundary must mostly depend on lattice effects.
It may well be possible that the plaquette order survives at $S>1$ and has a proximity effect near the RK point even at $S=1$. 
This could explain why numerical simulations of the square lattice dimer model struggle to establish its true ground state in this regime~\cite{Sachdev1989,Leung1996,Zeng1997,Syliuasen2006,Ralko2008,%
Banerjee2014,Oakes2018,HerzogArbeitman2019}. 
%
%

\section{Cubic lattice}
\label{sec:Basic-formulation_cubic}

The calculations for other lattices are straightforward generalisations 
of the square lattice case, with minimal but informative modifications. 
We begin with the cubic lattice, where we have the three 
lattice vectors $e_\eta=\hat{x},\hat{y},\hat{z}$. 
The corresponding RK Hamiltonian can be written as: 
\begin{eqnarray}
H_{D} &=& 
\sum_{r}\left\{ 
  -J\,b_{r,1}^{\dagger}b_{r+e_2,1}^{\dagger}
	    b_{r,2}^{\;}b_{r+e_1,2}^{\;}
	+\:(1 \leftrightarrow 2) 
\right.
\nonumber \\*
&& \qquad 
\left.
  + V\,b_{r,1}^{\dagger}b_{r,1}^{\;}
       b_{r+e_2,1}^{\dagger}b_{r+e_2,1}^{\;}
  +\:(1 \leftrightarrow 2) 
\right\} 
\nonumber \\*
&+& (12) \leftrightarrow (13) \leftrightarrow (23) , 
\label{eq:RK_Hamiltonian-1}
\end{eqnarray}
subject to the equivalent large-$S$ constraint 
\begin{equation}
\Pi_{r}\equiv\sum_{l\in v_{r}}b_{l}^{\dagger}b_{l}^{\;}-S=0 .
\label{eq:Constraint_S cubic}
\end{equation}
Using the radial gauge expression~\eqref{eq:radial_representation} for the bosonic field with $z =6$, this constraint can be written as in~\eqref{eq:Cobstraint_solution}, which in Fourier space reduces to 
\begin{equation}
\sum_{\eta} c_\eta \delta\rho_{\eta}(k) = 0
\, . 
\label{eq:k_space_constraint-1}
\end{equation}
Contrary to the case of the square lattice, we have now three fields 
$\delta\rho_\eta$ and one constraint, leaving two independent 
field variables. 
The Hamiltonian is made of three terms equivalent to the square lattice 
Eq.~\eqref{eq:RK_Hamiltonian_radial}, upon replacing 
$(12) \leftrightarrow (13) \leftrightarrow (23)$. 

As we did for the square lattice QDM, we derive here the resulting field 
theory to quadratic order. The calculation is similar to the one carried out 
in Sec.~\ref{sec:Fluctuations1}, keeping the constraints implicit. 
It is also possible to explicitly resolve the constraints and obtain an 
action in terms of two independent real scalar fields, as in 
Sec.~\ref{sec:Fluctuations2}, but for brevity, we omit the details of the calculation
and only present the final result at the end of this section. 

The terms in the cubic 
Hamiltonian are equivalent to combining the square lattice terms for the 
$(12)$, $(13)$, and $(23)$ components, see 
Eqs.~\eqref{eq: rho cos expansion square},~\eqref{eq: rho expansion J}, 
and~\eqref{eq: rho expansion V}, up to a factor of $4/9$ due to the fraction 
$S/6$ replacing $S/4$ in Eq.~\eqref{eq:radial_representation}: 
\begin{widetext}
\begin{equation}
\mathcal{S} = \int \!\!\ud\tau \: 
\left\{ 
  \sum_{k,\mu} i \delta\rho_\mu(k) \partial_\tau \Phi_\mu(-k) 
	+ 
	\frac{J S^2}{18} 
	\sum_{r,\mu} 
    \Big[
		  1 - \cos(\phi_\mu(r,\tau))
		\Big]
	+ 
	\sum_{k,\mu,\nu} 
		\frac{\mathcal{D}_{\mu\nu}}2 \delta\rho_\mu(k) \delta\rho_\nu(-k) 
\right\}  , 
\label{eq: cubic cosine action}
\end{equation}
where 
\begin{equation}
\mathcal{D} = 2
\left(
\begin{array}{ccc}
2(J-V) + (2V-J)(c_2^2+c_3^2) & -J c_1 c_2 & -J c_1 c_3
\\ 
-J c_1 c_2 & 2(J-V) + (2V-J)(c_1^2+c_3^2) & -J c_2 c_3 
\\ 
-J c_1 c_3 & -J c_2 c_3 &  2(J-V) + (2V-J)(c_1^2+c_2^2)
\end{array}
\right) ,
\label{eq:cubic matrix D}
\end{equation}
\end{widetext}
and we labelled $\phi_\mu$ the argument of the cosine term involving the 
phase fields $\Phi_\nu$ with $\nu\neq\mu$. 
%
%

\subsection{Action without instantons}
\label{subsec:Action-without-instantons cubic lattice}

When we expand about one given minimum,  
\begin{equation}
\frac{JS^{2}}{18}\sum_{\mu=1,2,3}
  [1-\cos\left(\phi_\mu\left(r,\tau\right)\right)]
\simeq
\frac{JS^{2}}{36}\sum_{\mu=1,2,3}\phi_\mu\left(r,\tau\right)^{2},
\label{eq:Taylor_no_instantons cubic}
\end{equation}
we see that integrating out the fields $\phi_\mu$ requires some care, since 
they are not all independent of one another. 

Following the same steps as for the square lattice dimer model, it is 
convenient to introduce in Fourier space the fields 
\begin{eqnarray}
\phi_1(k) &\equiv& e^{ik(e_2+e_3)/2} \, \tilde\phi_1(k) 
= 
2 \left[ 
  c_3 \Phi_2(k) - c_2 \Phi_3(k) 
\right]
\nonumber \\*
\phi_2(k) &\equiv& e^{ik(e_3+e_1)/2} \, \tilde\phi_2(k) 
= 
2 \left[ 
  c_1 \Phi_3(k) - c_3 \Phi_1(k) 
\right]
\nonumber \\*
\phi_3(k) &\equiv& e^{ik(e_1+e_2)/2} \, \tilde\phi_3(k) 
= 
2 \left[ 
  c_2 \Phi_1(k) - c_1 \Phi_2(k) 
\right]
\, . 
\nonumber \\
\label{eq: phi(k) cubic lattice}
\end{eqnarray}
These are most conveniently expressed as $\phi_\mu = \mathcal{Z}_{\mu\nu} \Phi_\nu$, 
where $\mathcal{Z}_{\mu\nu} = 2 \varepsilon_{\mu\nu\lambda} c_\lambda$ 
($\varepsilon_{\mu\nu\lambda}$ is the fully antisymmetric tensor 
and summation is implied). 

Notice that $\mathcal{Z}$ is a nonzero traceless antisymmetric matrix. It has one 
zero eigenvalue, $\sum_\nu \mathcal{Z}_{\mu\nu} c_\nu = 0$, and the other two must be 
non-vanishing and opposite, $\pm \zeta$: 
\begin{align}
2 \zeta^2 &= {\rm tr} \mathcal{Z}^2 = -2(c_1^2+c_2^2+c_3^2),
&
\zeta &= i \sqrt{c_1^2+c_2^2+c_3^2}.
\end{align}
The two non-vanishing eigenvectors define the physical space, and the null 
one is the gauge degree of freedom. One can therefore construct a projector 
onto the physical space as $-\mathcal{Z}^2 / (c_1^2+c_2^2+c_3^2)$. 

The cosine term in the Hamiltonian reduces to 
\begin{eqnarray}
\frac{JS^{2}}{36} 
\sum_{\mu} 
\phi_\mu(k)\phi_\mu(-k)
&=& 
\frac{JS^{2}}{36} 
\phi^{T}(k)\phi(-k) 
\\ 
&=& 
\frac{JS^{2}}{36} 
\Phi^{T}(k) \mathcal{Z}^{T}(k) \mathcal{Z}(-k) \Phi(-k) 
\nonumber \\ 
&=& 
- \frac{JS^{2}}{36} 
\Phi^{T}(k) \mathcal{Z}^{2}(k) \Phi(-k) 
\nonumber \\ 
&=& 
\sum_{\mu\nu} \frac{\mathcal{M}_{\mu\nu}}2 \Phi_\mu(k)\Phi_\nu(-k) 
\, , 
\nonumber 
\end{eqnarray}
where we used the fact that $\mathcal{Z}(-k)=\mathcal{Z}(k)$ and $\mathcal{Z}^{T}=-\mathcal{Z}$, and we defined
\begin{eqnarray}
\mathcal{M} 
&=& 
- \frac{JS^{2}}{18} \mathcal{Z}^{2}
\nonumber
\\ 
&=& 
\frac{2JS^{2}}{9}
\left(
  \begin{array}{ccc}
    c_2^2+c_3^2 & -c_1 c_2 & -c_1 c_3
		\\ 
    -c_1 c_2 & c_1^2+c_3^2 & -c_2 c_3
		\\ 
    -c_1 c_3 & -c_2 c_3 & c_1^2+c_2^2
  \end{array}
\right)
\, . 
\end{eqnarray}
Integrating out the fields $\Phi_\mu$ then gives 
\begin{eqnarray}
\frac{1}{2} 
\sum_{\mu,\nu} 
\mathcal{M}^{-1}_{\mu\nu} 
\partial_\tau\delta\rho_\mu(k)
\partial_\tau\delta\rho_\nu(-k),
\label{eq:M}
\end{eqnarray}
so we can write the full quadratic action without instantons as
\begin{equation}
\Sa = \frac12 \int\!\!\ud\tau \sum_{k,\mu,\nu}
\Big[
  \mathcal{M}^{-1}_{\mu\nu}\partial_\tau\delta\rho_\mu(k)\partial_\tau\delta\rho_\nu(-k) +
  \mathcal{D}_{\mu\nu} \delta\rho_\mu(k) \delta\rho_\nu(-k)
\Big].
\label{eq: gaussian action cubic}
\end{equation}
The dispersion can be obtained from the eigenvalues of 
$ \mathcal{M}\mathcal{D} = - 2J S^2 \mathcal{Z}^2 \mathcal{D} / 9$, after projecting 
out the unphysical modes that do not satisfy the constraint 
$\sum_\mu c_\mu \delta\rho_\mu = 0$. This could be done formally by adding 
an infinite Lagrange multiplier, but in fact there is no need to do so because 
the only unphysical mode is trivially the zero mode of $\mathcal{Z}^2 \mathcal{D}$ 
-- as we had previously observed in the square lattice QDM. 
The two non-vanishing eigenvalues are 
\begin{eqnarray}
&&
\omega^2 = 
\frac{4JS^2}{9} 
\Big\{\vphantom{\sqrt{c_1^4}}
  J \left[\vphantom{\Big[} 
	  2(c_1^2 + c_2^2 + c_3^2) - (c_1^4 + c_2^4 + c_3^4)
	\right] 
\label{eq: w2 dispersion cubic}
\\* 
&& \; 
	+ 2 V \left[ \vphantom{\Big[}
	  c_1^4 + c_2^4 + c_3^4 - (c_1^2 + c_2^2 + c_3^2) 
	  + c_1^2 c_2^2 + c_1^2 c_3^2 + c_2^2 c_3^2 
		\right] 
\nonumber \\* 
&& \; 
  \pm 
	2 \vert J-V \vert 
	\sqrt{c_1^4 c_2^4 + c_1^4 c_3^4 + c_2^4 c_3^4 
	  -c_1^2 c_2^2 c_3^2 (c_1^2 + c_2^2 + c_3^2)}
\vphantom{\Big[}
\Big\} .
\nonumber 
\end{eqnarray}
\begin{figure}
	\includegraphics{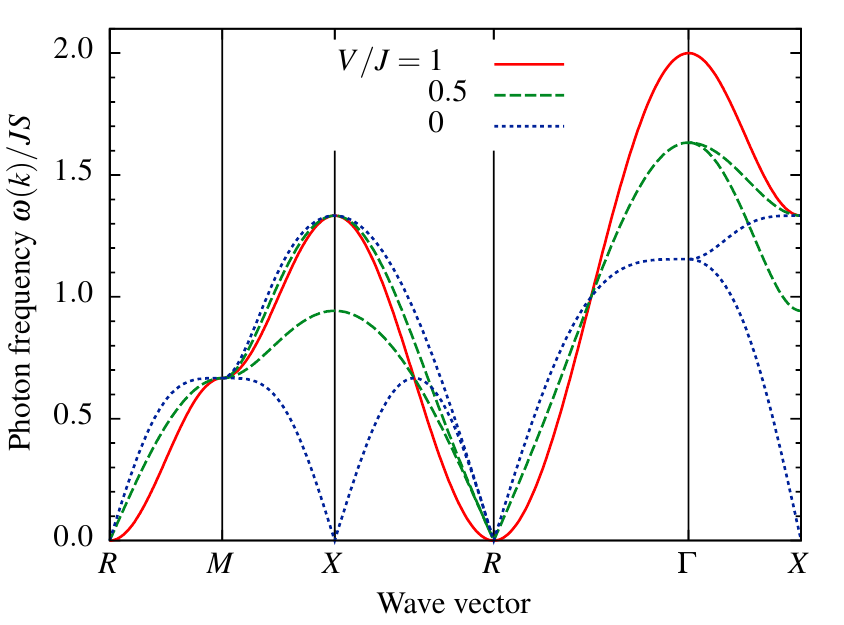}
	\caption{Photon dispersion relation of the large-$S$ QDM on the cubic lattice. The spectrum is gapless at the $R=(\pi,\pi,\pi)$ point; near this point, the dispersion is quadratic at the RK point and linear away from it. The spectrum has two non-degenerate branches away from the RK point. The lower branch develops minima at the $X=(\pi,0,0)$ points for lower values of $V$, which drives an ordering transition at $V=0$.}
	\label{fig:cubic dispersion}
\end{figure}
This dispersion is plotted for three values of $V/J$ in Fig.~\ref{fig:cubic dispersion}.
It is interesting to note that 
$\omega^2\vert_{J=V} \propto (c_1^2+c_2^2+c_3^2)^2$ 
and the two bands are degenerate at the RK point, with vanishing minima at 
$(\pi,\pi,\pi)$ and symmetry related points, and quadratic dispersion 
around them. 
Expanding near such minima, we find 
\begin{equation}
\omega^2 \simeq 
\frac{2S^2}{9} J(J-V) k^2 ,
\end{equation}
where $k$ is the (small) vector distance from the minimum, giving a 
speed of light 
\begin{equation}
c = \frac{S}{3} \sqrt{2J(J-V)} .
\label{eq:Speed_light}
\end{equation}

As mentioned earlier, one could have alternatively resolved the constraint 
explicitly, writing the three fields $\delta\rho_\mu$ in terms of two 
independent real scalar fields $h_a$ and resolving the corresponding 
inter dependence of the three fields $\phi_\mu$: 
\begin{equation}
c_1 \phi_1 + c_2 \phi_2 + c_3 \phi_3 = 0
\, . 
\label{dependence}
\end{equation}
For instance, one can do so via the 
relation $\phi_1 = -(c_2/c_1) \phi_2 - (c_3/c_1) \phi_3$ and 
$\delta\rho_\mu = \sum_a R_{\mu a} h_a$ with 
\begin{equation}
R
= 
2 
\left(
  \begin{array}{cc}
    c_2 & -c_3 
		\\ 
    c_1 & 0 
		\\ 
    0 & c_1 
  \end{array}
\right)
\, . 
\end{equation}
After a few lines of algebra, one arrives at the action 
\begin{align}
	\Sa = \frac12 \int \!\ud\tau \sum_{k,a,b} \Big[
	  \widetilde{\mathcal{M}}^{-1}_{ab} \partial_\tau h_a(k) \partial_\tau h_b(-k) +
	  \widetilde{\mathcal{D}}_{ab} h_a(k) h_b(-k) \Big]
\label{eq: cubic action h}
\end{align}
where $\widetilde{\mathcal{D}} = R^T \mathcal{D} R$ and 
\begin{equation*}
\widetilde{\mathcal{M}} =
\frac{JS^{2}}{18 c_1^2}
\left(
  \begin{array}{cc}
    c_1^2+c_2^2 & c_2 c_3 
		\\ 
    c_2 c_3 & c_1^1+c_3^2   \end{array}
\right).
\end{equation*}
One can easily verify that the action~\eqref{eq: cubic action h} gives indeed 
the same dispersion as Eq.~\eqref{eq: w2 dispersion cubic}. 
%
%

\subsection{Instantons}
\label{sec:Instantons cubic lattice}

The instanton contributions are calculated similarly to the square lattice case, by performing the Villain transformation \eqref{eq:Villain} on $\phi_\mu$ in \eqref{eq: cubic cosine action} and integrating out the smooth fields $\phi_\mu, h_\mu$. The result is a 3D Coulomb gas action for the integer charges $q_\mu$,
\begin{equation}
\mathcal{S} = \frac{(2\pi)^2}{2} \sum_{\omega,k} 
\sum_{\mu,\nu} q_{\mu}(-\omega,-k)
\left[\frac{18\omega^2}{JS^2}\mathcal{M}^{-1} + \mathcal{D}\right]^{-1}_{\mu\nu} q_{\nu}(\omega,k), 
\end{equation}
with the standard unscreened long-range Coulomb interaction. However, the constraint \eqref{dependence} on the fields $\phi_\mu$ implies the equivalent constraint
\begin{equation}
c_1q_1 + c_2q_2 + c_3q_3 = 0
\label{eq: dependence of q}
\end{equation}
on the instanton configurations $q_\mu(\omega,k)$ allowed in the low-energy sector.
In $(3+1)$ dimensions, such instantons are irrelevant and can be safely neglected.
%
%

\subsection{Large-$S$ phase diagram}

\begin{figure}
	\includegraphics{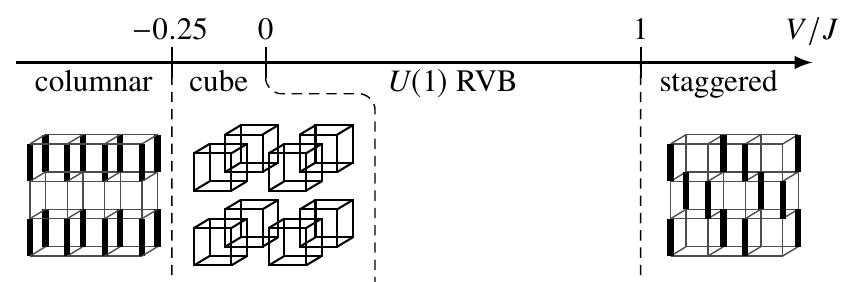}
	\caption{$S\to\infty$ phase diagram of the cubic lattice quantum dimer model. (To our knowledge, there are no conclusive studies of the $S=1$ QDM on the cubic lattice, to which one could compare the large $S$ results of this work.)}
	\label{fig:cubic phase diagram}
\end{figure}

Using the method described in Sec.~\ref{sec: square classical phase diagram}, we can obtain the ground state of the QDM Hamiltonian \eqref{eq:RK_Hamiltonian-1} in the limit $S\to\infty$. The results are summarised in Fig.~\ref{fig:cubic phase diagram}. 
Similarly to the square lattice case, we observe an extended $U(1)$ RVB liquid phase: in three dimensions, this phase is anticipated to survive at $S=1$. 
For $V>J$, the photon modes become unstable at the $(\pi,\pi,\pi)$ point, leading to staggered order. Likewise, the instability of the $(\pi,0,0)$ points for $V<0$ drives a transition into an RVB solid phase with isolated, resonating cubes. For $V<-J/4$, this phase gives way to columnar order. 

%
%

\section{Triangular lattice} 
\label{sec:triangular-lattice}

\begin{figure}
	\includegraphics{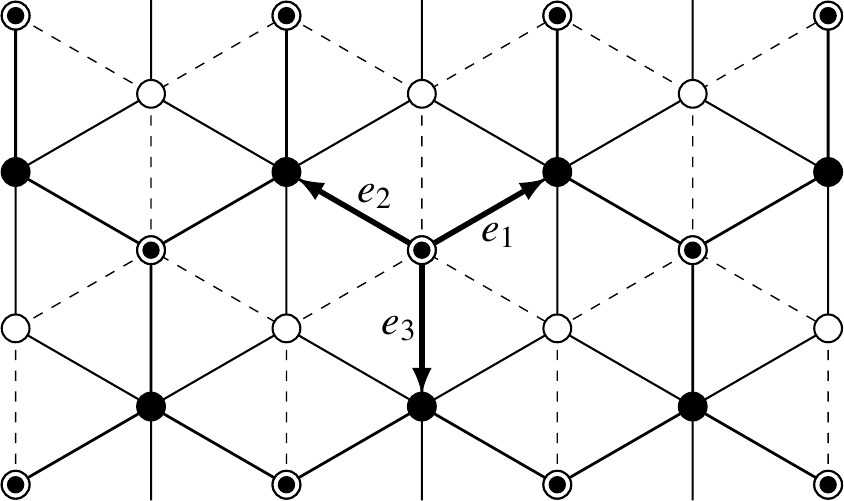} 
	\caption{\label{fig:triangular lattice}
		The triangular lattice illustrating a choice of base vectors $e_{1,2,3}$. 
		Its three sublattices are shown as solid dots, open circles, and dotted 
		circles. Some bonds of the lattice appear as dashed rather than solid lines, 
		in accordance with the correspondence to a projected cubic lattice 
		discussed  in the main text.}
\end{figure}

It is interesting to consider the case of the triangular lattice QDM 
immediately after the cubic one. It has sixfold connectivity and is tripartite, 
and one can view it as a cubic lattice projected along one of the [111] 
directions. 
Following the notation in Fig.~\ref{fig:triangular lattice}, the solid dots 
belong to one cubic 
sublattice, the open circles to the other, and the dotted circles are sites 
that belong to both cubic sublattices but get projected onto one another. 
The rhombic plaquettes of the triangular lattice correspond to the three 
independent faces of a cube, and therefore one can precisely identify flippable 
plaquettes and plaquette flipping terms between the two models. 
This projective view allows us to draw a complete correspondence between the 
two QDMs. 
Formally, the large-$S$ path integral approach presented in this paper proceeds 
identically, down to the numerical prefactors, for the triangular 
and cubic cases and we end up with the same two-component field 
theory~\cite{footnote_triplaq}, 
with action given (to quadratic order) by Eq.~\eqref{eq: gaussian action cubic} 
and dispersion given by Eq.~\eqref{eq: w2 dispersion cubic}. 
The correspondence holds only so long as we express the positions 
and wave vectors formally as $r$ and $k$, and the basis vectors as 
$e_1$, $e_2$ and $e_3$. 

To study the triangular lattice QDM, one then needs to substitute 
$k=(k_x,k_y)$ and a given choice of base vectors, for example 
$e_1=(\sqrt{3},1)/2$, $e_2=(-\sqrt{3},1)/2$, and $e_3=(0,-1)$, 
illustrated in Fig.~\ref{fig:triangular lattice}. 
This is, however, beyond the scope of the present paper. For $S=1$, we 
expect a $\mathbb{Z}_2$ liquid phase to be stable around the RK 
point~\cite{Moessner2001};
its fate, however, is unclear in the large-$S$ limit. 
If the description is able to capture it at all, it can only be after 
accounting for instantons and understanding how their role differs in 
bipartite and non-bipartite lattices -- a task which promises to pose 
non-trivial challenges. 

%
%

\section{Honeycomb lattice} 
\label{sec:Hexagonal-lattice}

\begin{figure}
	\includegraphics{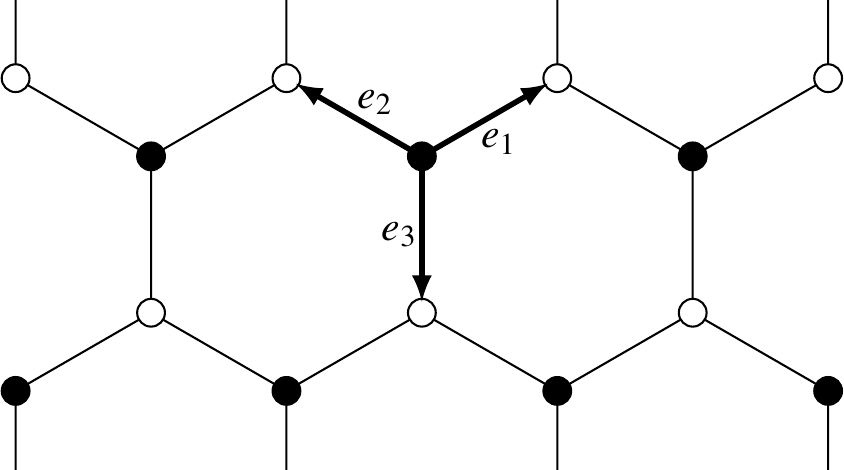} 
	\caption{Honeycomb lattice showing the lattice vectors $e_{1,2,3}$
		defined in the text.}
	\label{fig:honeycomb lattice}
\end{figure}

In this section, we consider the QDM on the honeycomb lattice, illustrated 
in Fig.~\ref{fig:honeycomb lattice}, and we present only the approach 
in which the constraint is resolved explicitly. 
Contrary to the cases considered so far, the primitive cell of the lattice 
contains two distinct sites (shown as solid dots and open circles in the figure). 
With the choice of lattice vectors 
$e_1=(\sqrt{3},1)/2$, $e_2=(-\sqrt{3},1)/2$, and 
$e_3=(0,-1)$ in Fig.~\ref{fig:honeycomb lattice}, 
we can write the bosonic representation of the Hamiltonian as 
\begin{align}
H_{D}= & 
\sum_{r \in A}\Big\{ 
  -J\,b_{r,1}^{\dagger}\,
	    b_{r+e_1-e_3,2}^{\dagger}\,
	    b_{r+e_2-e_3,3}^{\dagger}
\label{eq:RK_Hamiltonian honeycomb}
\\*
 & 
\qquad \quad \; \times 
	    b_{r+e_1-e_3,3}^{\;}\,
			b_{r+e_2-e_3,1}^{\;}\,
			b_{r,2}^{\;}
	+\:{\rm h.c.} 
\nonumber \\*
   + V & \,b_{r,1}^{\dagger}\,b_{r,1}^{\;}\,
	     b_{r+e_1-e_3,2}^{\dagger}\,b_{r+e_1-e_3,2}^{\;}\,
	     b_{r+e_2-e_3,3}^{\dagger}\,b_{r+e_2-e_3,3}^{\;}
\nonumber \\*
	 + V & \,b_{r+e_1-e_3,3}^{\dagger}\,b_{r+e_1-e_3,3}^{\;}\,
		    	b_{r+e_2-e_3,1}^{\dagger}\,b_{r+e_2-e_3,1}^{\;}\,
			    b_{r,2}^{\dagger}\,b_{r,2}^{\;}
\Big\} .
\nonumber
\end{align}

\begin{figure}
\includegraphics{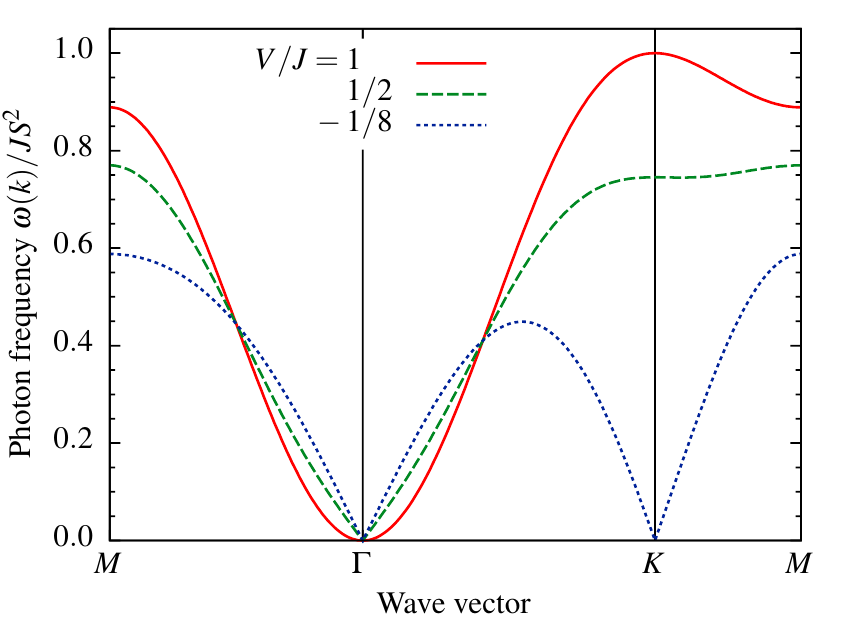}
\caption{Photon dispersion relation of the large-$S$ QDM on the honeycomb lattice. The spectrum is gapless at the $\Gamma$ point. Another minimum develops for small $V$ at the $K$ points; however, a plaquette ordering transition occurs before this minimum would become unstable.}
\label{fig:honeycomb dispersion}
\end{figure}

One has to write two separate (large-$S$) constraints, one for each 
sublattice: 
\begin{align}
\sum_{l \in v_{r \in A}}b_{l}^{\dagger}b_{l}^{\;}-S &= 0 ,
&
\sum_{l \in v_{r \in B}}b_{l}^{\dagger}b_{l}^{\;}-S &= 0 .
\label{eq:Constraint_S honeycomb}
\end{align}
The rest of the calculation follows rather straightforwardly from the square 
lattice case, barring some added algebraic difficulties, and is presented 
for completeness in App.~\ref{app: honeycomb details}. 
The argument of the cosine leads us to introduce the field 
\begin{align}
\phi(k) =
2 i \Big[ &
  \Phi_1(k) (s_2 c_3 - s_3 c_2) 
  + \Phi_2(k) (s_3 c_1 - s_1 c_3) 
\nonumber \\* 
&
  + \Phi_3(k) (s_1 c_2 - s_2 c_1) 
\Big] 
\nonumber 
\end{align}
and the following convenient resolution of the constraint in terms of a 
scalar field $h(k)$: 
\begin{eqnarray}
\delta\rho_1(k) &=& -2i (s_2 c_3 - s_3 c_2) \, h(k) 
\nonumber \\* 
\delta\rho_2(k) &=& -2i (s_3 c_1 - s_1 c_3) \, h(k) 
\nonumber \\* 
\delta\rho_3(k) &=& -2i (s_1 c_2 - s_2 c_1) \, h(k) 
. 
\end{eqnarray}
To quadratic order, one arrives then at the action 
\begin{eqnarray}
{\Sa} &=& \!\int \!\!\ud\tau \sum_r \: 
\Big\{
  i h(r,\tau)\partial_{\tau}\phi(r,\tau)
	+ \frac{2JS^{3}}{27} \left[ 1 - \cos(\phi(r,\tau)) \right] \!
\Big\} 
\nonumber\\*
&+& \int \!\!\ud\tau 
\sum_{k} \frac{D(k)}2 \: h\left(k,\tau\right)h\left(-k,\tau\right) ,
\label{eq: action h,phi honeycomb}
\end{eqnarray}
where 
\begin{eqnarray}
D(k) &=& 
\frac{8JS}{3} \big( s_{12}^4 +s_{23}^4 + s_{31}^4 \big)
\nonumber \\ 
&-&
\frac{16JS}{3} \big( 
s_{12}s_{23}c_{12}c_{23} + 
s_{23}s_{31}c_{23}c_{31} + 
s_{31}s_{12}c_{31}c_{12}
\big)
\nonumber \\ 
&+&
\frac{16VS}{3} \Big[
  s_{12}s_{23}(c_{12}c_{23}-s_{12}s_{23})
  \nonumber \\*
  && \qquad \, 
  +
  s_{23}s_{31}(c_{23}c_{31}-s_{23}s_{31})
  \nonumber \\*
  && \qquad \, 
  +
  s_{31}s_{12}(c_{31}c_{12}-s_{31}s_{12})
\Big],
\label{eq: D(k) honeycomb}
\end{eqnarray}
where we introduce for convenience
\begin{eqnarray}
s_{\mu\nu} &=& \sin[k(e_\mu-e_\nu)/2] = s_\mu c_\nu - s_\nu c_\mu
\nonumber\\*
c_{\mu\nu} &=& \cos[k(e_\mu-e_\nu)/2] = c_\mu c_\nu + s_\mu s_\nu.
\end{eqnarray}
Expanding about one given minimum,  
\begin{equation}
\frac{2JS^{3}}{27}[1-\cos\left(\phi\left(r,\tau\right)\right)]
\simeq
\frac{JS^{3}}{27}\phi\left(r,\tau\right)^{2},
\label{eq:Taylor_no_instantons honeycomb}
\end{equation}
and integrating over $\phi$, we arrive at the action 
\begin{align}
{\Sa} 
= \frac12
\int \!\ud\tau  \sum_{k} {}&
\Bigg[
  \frac{27}{2JS^{3}}\partial_{\tau}h(k,\tau)\partial_{\tau}h(-k,\tau)
\nonumber \\*
   & 
   +D(k) h(k,\tau)h(-k,\tau)
\Bigg],
\label{eq:General_form_h_honeycomb}
\end{align}
and the corresponding dispersion $\omega^2(k) = 2 J S^3 D(k) / 27$. 
This dispersion is plotted for three values of $V/J$ in Fig.~\ref{fig:honeycomb dispersion}.
It is gapless at the $\Gamma$ point for all values of $J$ and $V$. 
An instability develops for $V>J$ whereby $\omega^2$ becomes negative 
for small $k$. 
For $V \lesssim J$, the long-wavelength dispersion is linear: 
\begin{equation}
\omega \simeq \frac{\sqrt{2 J (J-V)}\, S^2}3  k.
\end{equation} 
As the value of $V$ is lowered, secondary minima appear at the $K$ points 
in the Brillouin zone which drive the system through 
an instability for $V < -J/8$.

Expanding $D(k)$ about its minimum at $\Gamma=\left(0,0\right)$, 
\begin{eqnarray}
D[(k_x,k_y)] &\simeq& 
  3S(J-V)(k_x^2 + k_y^2)
\nonumber \\* 
&+& 
	\frac{9S}{16}
	( 7 V - J ) (k_x^4 + k_y^4 + 2 k_x^2 k_y^2) ,
\nonumber 
\end{eqnarray}
we obtain the action: 
\begin{align}
{\Sa} =\frac12 \int \!\!\ud\tau \ud^2r \: &
\left[
  \frac{27}{2JS^{3}} (\partial_{\tau}h)^{2}
	+3S(J-V) (\nabla h)^{2}\right.
\nonumber \\*
&\left.+ \frac{9S}{16} (7 V - J) (\nabla^{2}h)^{2} 
\right]
+\dots
\label{eq:Action_2d_final honeycomb}
\end{align}
Contrary to the square lattice QDM, see Eq.~\eqref{eq:Action_2d_final}, 
we find that rotational symmetry is preserved near the band minimum 
of the honeycomb lattice QDM for $J \neq V$, at least up to quartic order. 
This is expected, since the discrete symmetry of the lattice upon 
$2\pi/3$ rotations implies that the first symmetry-breaking term allowed in 
the action must be of sixth order in $k$. 

\begin{figure}
	\includegraphics{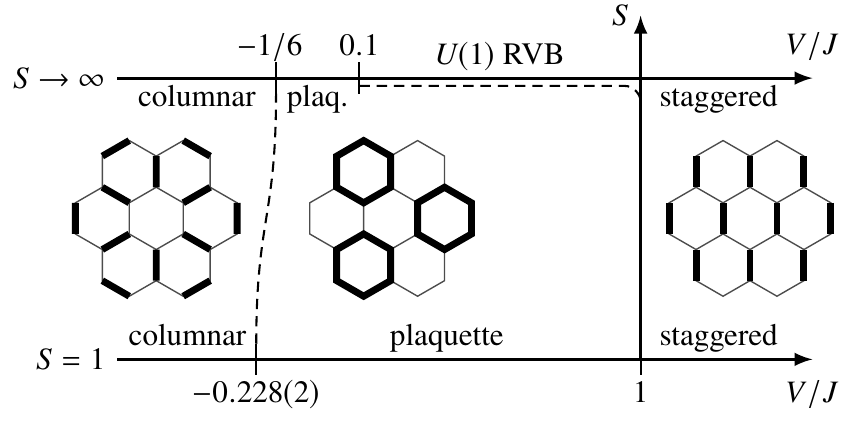}
	\caption{Phase diagram of the honeycomb lattice quantum dimer model at $S=\infty$ (this work) and $S=1$~\cite{Schlittler2017}.}
	\label{fig:honeycomb phase diagram}
\end{figure}

The ground state of the $S\to\infty$ model was also obtained as a function of $V/J$ by comparing the energy of several ordered and resonating phases suggested in the literature~\cite{Moessner2001honey,Schlittler2017} under the Hamiltonian \eqref{eq:RK_Hamiltonian honeycomb}. The resulting phase diagram is shown in Fig.~\ref{fig:honeycomb phase diagram}, together with the $S=1$ phase diagram obtained in Ref.~\cite{Schlittler2017}. 
As in the previous cases, the large-$S$ phase diagram predicts a columnar ordered, a plaquette ordered, an RVB liquid, and a staggered ordered phase, which are also sketched in Fig.~\ref{fig:honeycomb phase diagram}. 
It is interesting to note that the RVB liquid phase instability for $V<J$ occurs below the RVB--plaquette transition point (cf. Figs.~\ref{fig:honeycomb dispersion} and~\ref{fig:honeycomb phase diagram}); this suggests that the latter may in fact be a first order transition. 
At finite $S$, instantons are expected to gap out the $U(1)$ liquid phase in two dimensions, leading to its immediate collapse, see Sec.~\ref{sec:Instantons}. The other three phases are all observed at $S=1$ with critical $V/J$ similar to the large-$S$ result. 
%
%

\section{Diamond lattice}
\label{sec:Diamond-lattice}

\begin{figure}
	\includegraphics{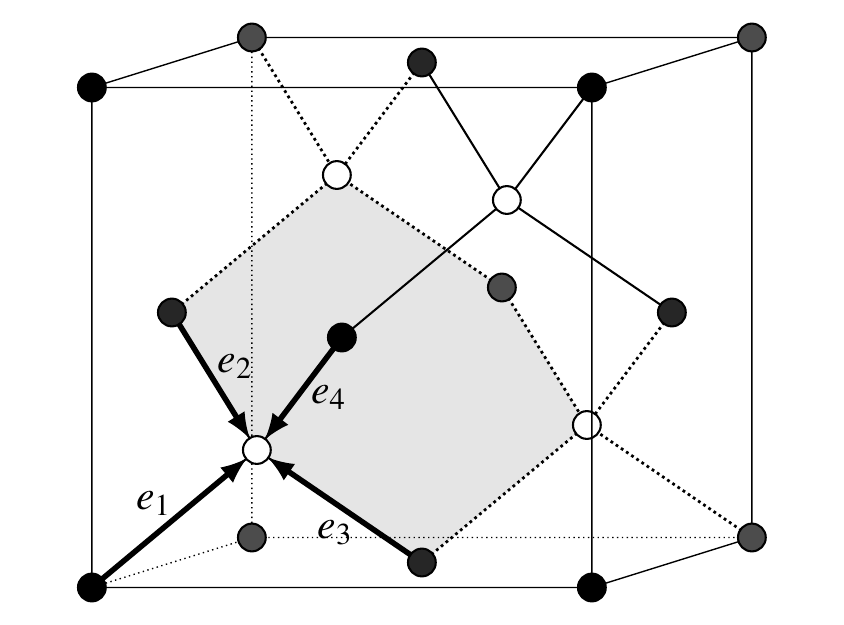} 
	\caption{Unit cell of the diamond lattice showing the lattice vectors $e_{1,\dots,4}$ introduced in the text. The two interpenetrating fcc lattices are shown as solid and open dots, respectively. A hexagonal plaquette is shown shaded by way of example.}
	\label{fig:diamond lattice}
\end{figure}

The diamond lattice is composed of two interpenetrating face
centred cubic (fcc) lattices, as shown in 
Fig.~\ref{fig:diamond lattice}. 
Sublattice $A$ is connected to sublattice $B$ by the vectors
\begin{align}
e_{1}&=\frac{1}{\sqrt{3}}\left(1,1,1\right) 
& 
e_{2}&=\frac{1}{\sqrt{3}}\left(1,-1,-1\right)
\nonumber\\
e_{3}&=\frac{1}{\sqrt{3}}\left(-1,1,-1\right) 
& 
e_{4}&=\frac{1}{\sqrt{3}}\left(-1,-1,1\right) ,
\label{eq:4_vectors}
\end{align}
which we chose to define the unit length in the system. 
Notice that $\sum_{\mu} e_\mu = 0$. 
The lattice is fourfold coordinated and the smallest lattice loop over which 
a dimer move can take place is a hexagon. Each hexagonal plaquette involves 
lattice bonds of three out of the four types, and therefore there are four 
inequivalent plaquette terms, involving respectively 
$\mu = 1,2,3$, $\mu = 1,2,4$, $\mu = 1,3,4$, and $\mu = 2,3,4$. 

The QDM Hamiltonian is the sum of four copies of the honeycomb lattice Hamiltonian~\eqref{eq:RK_Hamiltonian honeycomb} 
and we refrain from writing it here explicitly for convenience 
(see Sec.~\ref{sec:Hexagonal-lattice} and App.~\ref{app: honeycomb details}). 
Expressing the bosonic operators in the radial gauge  \eqref{eq:radial_representation} and expanding to quadratic order in $\delta\rho$ leads to the action
\begin{eqnarray}
{\Sa} &=& 
\int \!\!\ud\tau \sum_{k,\mu} \: 
  i \delta\rho_\mu(k,\tau)\partial_{\tau}\Phi_\mu(-k,\tau)
\nonumber \\*
&+& \int \!\!\ud\tau \sum_{r,\alpha} \: 
	\frac{JS^{2}}{8} \left[ 1 - \cos(\phi_\alpha(r,\tau)) \right] 
\nonumber \\*
&+& \int \!\!\ud\tau 
\sum_{k,\mu,\nu} \frac{{\Da}_{\mu\nu}(k)}2 \: 
\delta\rho_\mu\left(k,\tau\right)
\delta\rho_\nu\left(-k,\tau\right) ,
\label{eq: action drho,Phi diamond}
\end{eqnarray}
where $\alpha$ runs over the inequivalent plaquette sublattices, ${\Da}_{\mu\nu}(k)$ is the interaction matrix that follows from expanding the Hamiltonian terms $-2J \sqrt{\rho\rho\rho\rho\rho\rho}$ and $V \rho\rho\rho$ to quadratic order, and $\phi$ is the lattice curl of $\Phi$ around each plaquette.
In Fourier space, we have
\begin{equation}
\phi_\alpha(k) 
= 
2 i \sum_{\beta \gamma \delta} 
\varepsilon_{\alpha \beta \gamma \delta} \Phi_\beta(k) s_\gamma c_\delta,
\label{eq: phi_eta diamond}
\end{equation}
where $\varepsilon_{\alpha \beta \gamma \delta}$ is the totally 
antisymmetric tensor,  $c_\mu= \cos(k e_\mu/2)$,
and $s_\mu = \sin(k e_\mu/2)$. 

Since the two sublattices of the diamond lattice are inequivalent, there are two separate (large-$S$) constraints on $\rho$, one for each sublattice: 
\begin{align}
\sum_{l \in v_{r \in A}}b_{l}^{\dagger}b_{l}^{\;}-S &= 0, 
&
\sum_{l \in v_{r \in B}}b_{l}^{\dagger}b_{l}^{\;}-S &= 0 .
\label{eq:Constraint_S diamond}
\end{align}
In Fourier space, these can be written as
\begin{equation}
\sum_{\mu} c_\mu \, \delta\rho_\mu(k) = \sum_{\mu} s_\mu \, \delta\rho_\mu(k) = 0;
\label{eq: constraints delta rho diamond}
\end{equation}
since we have two constraints for the four fields $\delta\rho_\mu$, there are two independent fields left. 
Resolving the constraints explicitly, however, leads to rather intractable 
algebra; therefore, we only present the implicit approach of 
Sec.~\ref{sec:Fluctuations1}.

\begin{figure}
\includegraphics{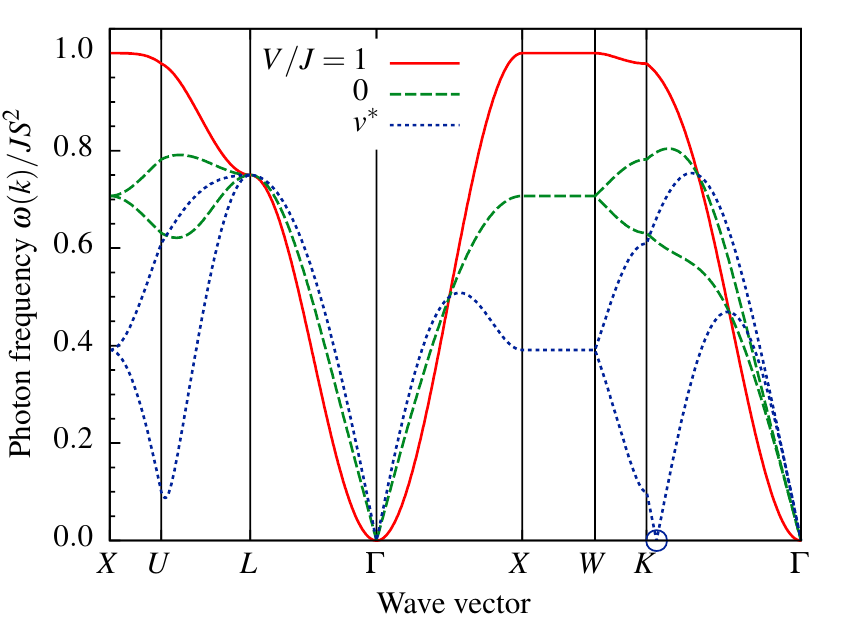}
\caption{Photon dispersion relation of the large-$S$ QDM on the diamond lattice.  The spectrum is gapless at the $\Gamma$ point and has two non-degenerate branches away from the RK point. Another minimum develops for small $V$ near the $K$ points; however, a plaquette ordering transition occurs before this minimum would become unstable at $V/J = v^*\approx-0.694$.}
\label{fig:diamond dispersion}
\end{figure}

Ignoring instantons (for the reasons discussed in Sec.~\ref{sec:Instantons cubic lattice}), the middle term of Eq.~\eqref{eq: action drho,Phi diamond} can be expanded about a given minimum of the cosine. 
After Fourier transforming and substituting \eqref{eq: phi_eta diamond}, we obtain
\begin{eqnarray}
{\Sa} &=& 
\int \!\!\ud\tau \sum_{k,\mu} \: 
  i \delta\rho_\mu(k,\tau)\partial_{\tau}\Phi_\mu(-k,\tau)
\nonumber \\*
&+& \int \!\!\ud\tau \sum_{k,\mu,\nu} \: 
	\frac{\mathcal{M}_{\mu\nu}}2 \Phi_\mu(k,\tau)\Phi_\nu(-k,\tau)
\nonumber \\*
&+& \int \!\!\ud\tau 
\sum_{k,\mu,\nu} \frac{{\Da}_{\mu\nu}(k)}2 \: 
\delta\rho_\mu\left(k,\tau\right)
\delta\rho_\nu\left(-k,\tau\right);
\label{eq: diamond quadratic action}
\end{eqnarray}
obtaining the matrices $\mathcal{M}$ and $\mathcal{D}$ from the QDM Hamiltonian is straightforward but algebraically tedious and is presented in App.~\ref{app: diamond details}. 

As in Sec.~\ref{sec:Fluctuations1}, the fields $\Phi_\mu$ can now be 
integrated out to obtain the action
\begin{align}
{\Sa} = \frac12 
\int \!\!\ud\tau \sum_{k,\mu,\nu} \: &
  \Big[ \left( \mathcal{M}^{-1} \right)_{\mu\nu}
	  \partial_{\tau}\delta\rho_\mu(k,\tau)
	  \partial_{\tau}\delta\rho_\nu(k,\tau)
\nonumber \\*
&+ {\Da}_{\mu\nu}(k) \: 
\delta\rho_\mu\left(k,\tau\right)
\delta\rho_\nu\left(-k,\tau\right) \Big].
\label{eq: action drho diamond}
\end{align}
The dispersion $\omega^2(k)$ is given by the eigenvalues of the matrix $\mathcal{MD}$. Two of these eigenvalues are identically zero: these correspond to the unphysical modes ruled out by the constraints \eqref{eq: constraints delta rho diamond}. 
The remaining two eigenvalues can be worked out by straightforward but rather lengthy algebra; the resulting dispersion is plotted along high-symmetry directions for three values of $V/J$ in Fig.~\ref{fig:diamond dispersion}. 

Away from the RK point, the resulting two photon modes are normally 
non-degenerate; near $k=0$, however, they agree to leading order: 
\begin{equation}
	\omega^2\simeq  \frac{S^4}6 J(J-V)k^2 ,
\end{equation}
giving a speed of light 
\begin{equation}
	c = S^2\sqrt{J(J-V)/6} .
\end{equation}
This can be contrasted to the quantum Monte Carlo result for $S=1$ 
in Refs.~\cite{Sikora2009,Sikora2011}, namely 
$c \simeq \sqrt{0.8 J(J-V)}$. 
We expect higher-order corrections in the 1/S expansion to substantialy reduce this discrepancy~\cite{Kwasigroch2017}.

\begin{figure}
	\includegraphics{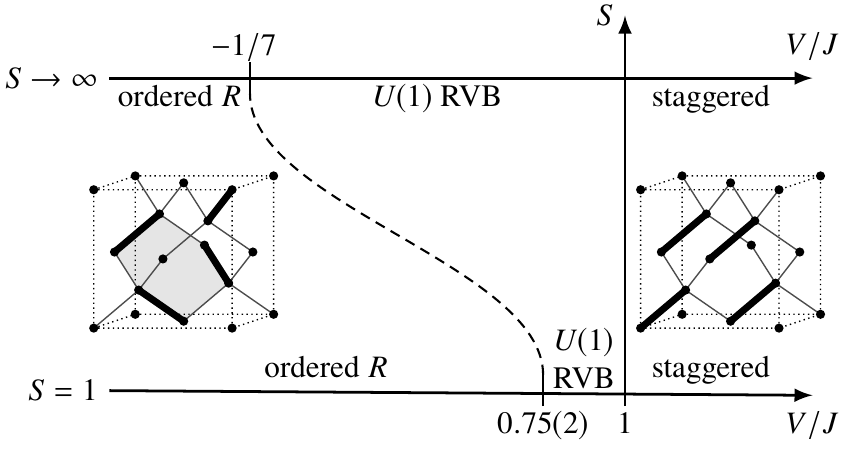}
	\caption{Phase diagram of the diamond lattice quantum dimer model at $S=\infty$ (this work) and $S=1$~\cite{Sikora2011}.}
	\label{fig:diamond phase diagram}
\end{figure}

The ground state of the $S\to\infty$ model was obtained as a function of $V/J$ by comparing the large-$S$ minimum energy of ordered and resonating phases suggested in the literature~\cite{Sikora2009,Sikora2011,Shannon2012}. 
The resulting phase diagram is shown in Fig.~\ref{fig:diamond phase diagram}, together with the $S=1$ phase diagram obtained in Ref.~\cite{Sikora2011}. 
Unlike the previous cases, no resonating solid phase is identified in either limit: both phase diagrams consist of staggered ordered, $U(1)$ liquid and ordered $R$~\cite{Sikora2009} phases. The liquid phase is far larger for $S\to\infty$ than at $S=1$: this is consistent with the intuition that soft dimers (and spins) favour fluctuating phases. 
%
%

\section{Relation to other field theoretic approaches}
\label{sec: other approaches}

For quantum dimers models on bipartite lattices, field-theoretic approaches were already known in the literature. These include height mappings in two dimensions~\cite{Fradkin_2013} and $U(1)$ gauge theory descriptions in three dimensions~\cite{Henley2010}. In this section, we briefly review them and discuss how they relate to our large-$S$ representation. For simplicity, we focus on the square and cubic lattices.
%
%

\subsection{Height mapping on the square lattice}
\label{app: 2d height mapping}

In $S=1$ dimer models, a height mapping can be constructed by assigning a height $\tilde{h}$ to all plaquettes of the lattice, starting from a reference plaquette with $\tilde{h}=0$. 
Going clockwise around a vertex in the $A$ sublattice (see Fig.~\ref{fig:square lattice}), the height field changes by $-1/4$ on crossing a bond without a dimer and by $+3/4$ on crossing one with a dimer. 
Likewise, going clockwise around a vertex in the $B$ sublattice, $\tilde{h}$ changes by $-3/4$ and  $+1/4$ on bonds with and without dimers, respectively. 
These numbers are the difference between the occupation of a given bond and the average occupation of all bonds, $1/4$. 
Therefore, this mapping can be generalised to the large-$S$ case by requiring that 
\begin{equation}
\Delta\tilde{h}=\pm\delta\rho,
\label{eq:height mapping 2d large-S}
\end{equation}
where the sign depends on the 
direction in which the bond is traversed, 
as described above.

In order to coarse-grain this height description, it is useful to define the so-called magnetic field representation,
\begin{equation}
\vec{B}(r) = \left( B_{x}, B_{y}, 0 \right), 
\quad
B_{\mu}(r)=\left(-1\right)^{x+y}\delta\rho_{\mu}(r),
\label{eq:Magnetic_field_2d}
\end{equation}
in which the height mapping \eqref{eq:height mapping 2d large-S} can be expressed as
\begin{equation}
B_\mu(e+r_\mu/2) = \tilde{h}[r+(e_\mu+\varepsilon_{\mu\nu}e_\nu)/2] - \tilde{h}[r+(e_\mu-\varepsilon_{\mu\nu}e_\nu)/2] .
\label{eq:height mapping magnetic field}
\end{equation}
In reciprocal space, the signs appearing in \eqref{eq:Magnetic_field_2d} have the effect of shifting the gapless point of the dispersion \eqref{eq: dispersion QDM square} from $(\pi,\pi)$ to $(0,0)$, that is, coarse-graining will capture the long-wavelength, low-frequency features of the gapless modes. In particular, \eqref{eq:height mapping magnetic field} can be Fourier transformed to give
\begin{equation}
B_\mu(k) = 2i\varepsilon_{\mu\nu} \sin(k_\nu/2) \tilde{h}(k);
\label{eq:height mapping reciprocal space}
\end{equation}
comparing with \eqref{eq:square lattice resolved constraint} immediately gives 
\begin{equation}
\tilde{h}(k) = h[k+(\pi,\pi)] \implies
\tilde{h}(r) = (-1)^{x+y} h(r),
\end{equation}
where $h$ is the single-component height field introduced in Sec.~\ref{sec:Fluctuations2}. This concludes the reconstruction of the height mapping for classical (nondynamical) dimer models.

To recover quantum dynamics, the conjugate variables $\Phi_\mu$ and $\phi$ can be reexpressed as
\begin{eqnarray}
\vec{\Phi}(r) = \left( \tilde\Phi_{x}, \tilde\Phi_{y}, 0 \right), 
\quad
\tilde\Phi_{\mu}(r)&=&\left(-1\right)^{x+y}\Phi_{\mu}(r)\\
\tilde\phi(r) &=& \left(-1\right)^{x+y}\phi(r),
\end{eqnarray}
which leads to Berry phase terms of the form
\begin{eqnarray}
i \sum_{k,\mu} \delta\rho_\mu(k)\partial_\tau \Phi_\mu(-k) &=& i\sum_k \vec{B}(k)\cdot\partial_\tau \vec\Phi(-k)\\
i\sum_k h(k)\partial_\tau\phi(-k) &=& i\sum_k \tilde{h}(k)\partial_\tau\tilde{\phi}(-k).
\end{eqnarray}
Integrating $\tilde\phi$ out and coarse-graining (that is, focusing on small $k$) then yields the quadratic action \eqref{eq:Action_2d_final}.
%
%

\subsection{Coulomb gauge theory on the cubic lattice}
The key idea of the mapping discussed above is turning the lattice gauge field and its divergence-free condition \eqref{eq:Constraint_S} into a coarse-grained, true vector field with $\nabla\cdot\vec{B} = 0$. Indeed, for small $k$, \eqref{eq:height mapping reciprocal space} reduces to 
\begin{equation}
\vec{B}(k) = ik\times \vec{h}(k) \quad \Longleftrightarrow \quad \vec{B} = \nabla\times\vec{h},
\end{equation}
where $\vec{h} = (0,0,\tilde{h})$. Similarly, we now want to construct a divergence-free magnetic field $\vec{B}$ on the cubic lattice by coarse-graining $\delta\rho_\mu$. To achieve this, we express our field in terms of a magnetic vector potential: $\vec{B}=\nabla\times\vec{A}$.

The construction again starts by defining
\begin{equation}
B_{\mu}(r) = \left(-1\right)^{x+y+z}\delta\rho_\mu (r);
\label{eq:Magnetic_field 3d}
\end{equation}
similarly to the square lattice case, this amounts to shifting the gapless point of the photon dispersion from $(\pi,\pi,\pi)$ to the origin.
The reciprocal space constraint \eqref{eq:k_space_constraint-1} now becomes 
\begin{equation}
\sum_\mu 2i\sin(k_\mu/2)B_\mu(k) = 0.
\end{equation}
This constraint can be resolved similarly to \eqref{eq:height mapping reciprocal space} by introducing another vector field $\vec{A}$:
\begin{equation}
B_\mu(k) = 2i\varepsilon_{\mu\nu\lambda} \sin(k_\nu/2) A_\lambda(k);
\label{eq:height mapping reciprocal space 3D}
\end{equation}
at small $k$, this reduces to $B=\nabla\times A$, as desired.

We now turn the conjugate variables $\Phi$ and $\phi$ into vector fields by writing
\begin{eqnarray}
\vec\Phi_\mu(r) &=&  (-1)^{x+y+z}\Phi_\mu(r)\\
\vec\phi_\mu(r) &=&  (-1)^{x+y+z}\phi_\mu(r).
\end{eqnarray}
Eq.~\eqref{eq: phi(k) cubic lattice} can be written as
\begin{equation}
\vec\phi_\mu(k) = 2i\varepsilon_{\mu\nu\lambda} \sin(k_\nu/2) \vec\Phi_\lambda(k)
\end{equation}
and the Berry phase can be integrated by parts to obtain
\begin{equation}
i\sum_{k,\mu} \delta\rho_\mu(k)\partial_\tau\Phi_\mu(-k)
=
i\sum_k \partial_\tau\vec{A}(k) \cdot\vec{\phi}(-k).
\end{equation}
Expanding to quadratic order in $\phi$ around a given minimum of the cosine, we finally obtain the action
\begin{eqnarray}
\Sa = \int\ud\tau\sum_k&&\Big[ i\partial_\tau\vec{A}(k) \cdot\vec{\phi}(-k) + \frac{JS^2}{36} \vec\phi(k)\cdot\vec\phi(-k)\nonumber\\*
&&+\frac{\widetilde{\mathcal{D}}_{\mu\nu}}2 A_\mu(k)A_\nu(-k)\Big],
\end{eqnarray}
where $\widetilde{\mathcal{D}}$ is given in terms of the matrix $\Da$ defined in \eqref{eq:cubic matrix D} by
\begin{equation}
\widetilde{\Da}_{\mu\nu}(k) = 4\varepsilon_{\mu\kappa\rho}\varepsilon_{\nu\lambda\sigma} \sin(k_\kappa/2)\sin(k_\lambda/2) \Da_{\rho\sigma}[k+(\pi,\pi,\pi)].
\end{equation}
Finally, we integrate out $\phi$ and expand $\widetilde{\Da}$ around $q=0$ to obtain a coarse-grained action in terms of $A$ only. Away from the RK point, the leading order terms give
\begin{equation}
\Sa \simeq \int\!\!\ud\tau\sum_k\bigg[\frac9{JS^2} \partial_\tau\vec{A}(k)\cdot\partial_\tau\vec{A}(-k) 
+ 2(J-V) \vec{B}(k)\cdot\vec{B}(-k)\bigg].
\label{eq:cubic Coulomb final}
\end{equation}
Similarly to ordinary quantum electrodynamics, we can identify $\partial_\tau\vec{A}$ as the electric field; that is, \eqref{eq:cubic Coulomb final} is the action of a linearly dispersing $U(1)$ gauge theory. The speed of light is given by
\begin{equation}
c=\frac{S}{3}\sqrt{2\left(J-V\right)J}, 
\label{eq:Speed_light Coulomb}
\end{equation}
in agreement with \eqref{eq:Speed_light}. 
At the RK point, the $B^2$ term vanishes; to leading order, the action becomes
\begin{eqnarray}
\Sa \simeq \int\ud\tau\sum_k&&\bigg\{\frac9{JS^2} \partial_\tau\vec{A}(k)\cdot\partial_\tau\vec{A}(-k) 
\label{eq:cubic Coulomb final expanded}
\\*
&&-\frac{J}2 \big[k\times\vec{B}(k)\big]\cdot\big[k\times\vec{B}(-k)\big]\bigg\}.
\nonumber
\end{eqnarray}
%
%

\section{Conclusion}
\label{sec:Conclusions}
We proposed a general route to obtain the field-theoretic action for 
microscopic Hamiltonians with hard constraints, based on a slave boson 
representation of the relevant degrees of freedom and their constraints, 
combined with a large-$S$ path integral formulation. 

We used it to systematically derive Lagrangians for bipartite QDMs in 
2D and 3D from the corresponding microscopic Hamiltonians. 
We find good agreement with known results in the literature; namely, 
calculations up to quadratic order yield a stiffness for the square lattice 
QDM at the RK point equal to $1/4$ compared to the exact result 
$\pi/18$~\cite{Moessner2011,Tang2011}; and they yield a 
speed of light in the gapless phase of the diamond lattice QDM 
$c=\sqrt{J(J-V)/6}\,\,S^2$ compared to the numerical result 
$c(S=1) \simeq \sqrt{0.8 J(J-V)}$~\cite{Sikora2009,Sikora2011}. 

Our approach applies straightforwardly to the 
nonbipartite case of the QDM on the triangular lattice, where we observe an 
intriguing analytical relation to the formalism of the cubic 
lattice, which will be interesting to explore in future work. 
%
%

\section*{Acknowledgements}

We thank Baptiste Bermond, John Chalker, and Roderich Moessner for useful discussions. 
A.M.T.\ is supported by the Condensed Matter Physics and Materials
Science Division, in turn funded by the U.S. Department of Energy,
Office of Basic Energy Sciences, under Contract No. DE-SC0012704.
G.G.\  was supported by the U.S Department of Energy, Office of Science, Basic Energy Sciences as a part of the Computational Materials Science Program.
This work was supported, in part, by the Engineering and Physical
Sciences Research Council (EPSRC) Grant No. EP/M007065/1 (C.C.\ and
G.G.) and Grant No. EP/P034616/1 (C.C.). 
Statement of compliance with the EPSRC policy framework on
research data: this publication reports theoretical work that does
not require supporting research data. 
%
%
\appendix

\section{Instanton measure on the square lattice}
\label{app: instanton}

Instantons appear naturally in the compact gauge theory as stationary trajectories of the action with $\phi$ changing by $2\pi$ on a given site between $\tau=\pm\infty$. Unlike the point-like instantons described in Sec.~\ref{sec:Instantons}, these objects are smooth as a function of time, and thus have a nontrivial instanton core. The action can then be expanded to quadratic order around such solutions, similarly to the case $h=\phi\equiv0$ shown in Sec.~\ref{subsec:Action-without-instantons}. The resulting fluctuation determinant will appear in the probability of instantons as a preexponential factor.

We thus have to construct such a stationary instanton solution. Using (\ref{eq: action h,phi}, \ref{eq:Villain}) with a single instanton event at $r=\tau=0$, we get the following quadratic action in terms of Fourier components:
\begin{align}
	\Sa = \int (\ud\omega)(\ud^2 k)\: &\bigg\{ \omega h(k,\omega)\phi(-k,-\omega)  \nonumber\\*
	&+ \frac{\Da_0(k)}{2} h(k,\omega)h(-k,-\omega) \nonumber\\*
	&+\frac{M}2 \left[ \phi(k,\omega) - \frac{2\pi i}\omega\right]\times\mathrm{c.c}\bigg\},
\end{align}
where we introduce $M=JS^2/8$ for brevity and $i/\omega$ is the Fourier transform of the Heaviside function in $\tau$. Since the different Fourier components are decoupled in this action, we can minimise with respect to them separately, resulting in the stationary action
\begin{align}
	h_0(k,\omega) &= \frac{2\pi i qM}{\omega^2 + M\Da_0(k)} \nonumber\\
	\phi_0(k,\omega) &= \frac{2\pi iq}\omega \frac{M\Da_0(k)}{\omega^2 + M\Da_0(k)}.
	\label{eq: stationary solution}
\end{align}
In real space, this corresponds to a point-like instanton described in Sec.~\ref{sec:Instantons} together with a power-law decaying ``instanton core.'' We should also note that \eqref{eq: stationary solution} is not a stationary trajectory under the original action, but it becomes one if the $M(1-\cos\phi)$ potential term is replaced with a continued parabolic potential, $V(\phi) = \frac{M}2 \min_n(\phi-2\pi n)^2$: Indeed, the only point where $\phi_0$ reaches $\pi$ is $r=\tau=0$, where the cusp in the potential is recovered by the external instanton charge. In the following, we will thus use $V(\phi)$.

The action can now be expanded to quadratic order in $\delta h$  and $\delta\phi$ around both the trivial trajectory $h=\phi\equiv0$ and the instanton trajectory \eqref{eq: stationary solution}. $\delta h$ can easily be integrated out in both cases, giving the following actions in $\delta \phi$:
\begin{align}
	\delta\Sa_0 &= \frac12 \int (\ud\omega)(\ud^2 k)
		\left(\frac{\omega^2}{\Da_0(k)}+M\right) \delta\phi(k,\omega)\delta\phi(-k,-\omega) 
		\label{eq:no-instanton action k omega, phi q}\\
	\delta\Sa_\mathrm{i} &= \delta\Sa_0 -\frac{\pi M}{\partial_\tau\phi_0(r=\tau=0)} [\delta\phi(r=\tau=0)]^2,
		\label{eq:instanton action k omega, phi q}
\end{align}
where the additional term in \eqref{eq:instanton action k omega, phi q} corresponds to the cusp of the continued parabolic potential at $\phi=\pi$, only reached at $r=\tau=0$. The most important difference between the two actions is that $\delta\Sa_\mathrm{i}$ has the zero mode $\psi = \partial_\tau\phi_0$, corresponding to the continuous time-translation symmetry of the setup. For such modes, the usual contribution to the partition function, $e^{-\Sa_\mathrm{cl}} (\det K)^{-1/2}$, is replaced by
\begin{equation}
	\ud\tau \sqrt{\frac{\langle \psi | \psi\rangle}{2\pi}} e^{-\Sa_\mathrm{cl}} (\det \tilde{K})^{-1/2},
\end{equation}
where $\Sa_\mathrm{cl}$ is the action due to the stationary instanton, given by \eqref{eq:single instanton action}, 
and $\tilde{K}$ is the fluctuation kernel of \eqref{eq:instanton action k omega, phi q} restricted to the non-zero modes. 
Suppose $|\psi\rangle$ is proportional to a basis vector (this can always be achieved using a unitary transformation on the kernel $K$). Then, $\tilde{K}$ is the principal minor of $K$ that excludes the row and column of $\psi$. $\det\tilde{K}$ is thus a cofactor of the full kernel, and we have
\begin{equation}
	\det\tilde{K} = \langle \tilde\psi | K^{-1} | \tilde\psi\rangle \det K
	\label{eq:cofactor formula}
\end{equation}
by the cofactor formula for matrix inversion, where $|\tilde\psi\rangle$ is the normalised zero mode $|\psi\rangle / \sqrt{\langle \psi|\psi\rangle}$.

In a matrix language, $K$ is the sum of the kernel $K_0$ appearing in \eqref{eq:no-instanton action k omega, phi q} and a dyad $-\lambda|v\rangle\langle v|$, where $|v\rangle$ corresponds to the $\delta$-function at $r=\tau=0$ implied in \eqref{eq:instanton action k omega, phi q} in an arbitrary basis. For such a matrix, we have the following:

	(i) $\det K = \det K_0 (1-\lambda \langle v|K_0^{-1}|v\rangle)$.

	(ii) If $1-\lambda \langle v|K_0^{-1}|v\rangle=0$, $K_0^{-1}|v\rangle$ is an eigenvector of $K$ with eigenvalue 0.
	
	(iii) $K^{-1} = K_0^{-1} + \dfrac{\lambda K_0^{-1} |v\rangle\langle v| K_0^{-1} }{1-\lambda \langle v| K_0^{-1}| v\rangle}.$
	
The first statement can be shown by inserting factors of $K_0^{1/2}K_0^{-1/2}$ into the definition of $K$; the other two are straightforward to verify. Now, $\det \tilde{K}$ follows as
\begin{align}
	\det\tilde{K} 
	&= \left\langle\tilde\psi_0\middle| 
			\left( K_0^{-1} + \frac{\lambda K_0^{-1} |v\rangle\langle v| K_0^{-1}}{1-\lambda \langle v| K_0^{-1}| v\rangle} \right)
			\middle|\tilde\psi_0\right\rangle \times \nonumber\\*
		&\qquad (1-\lambda \langle v| K_0^{-1}| v\rangle)\det K_0 
		\label{eq:instanton matrix algebra 1}\\
	&= \langle\tilde\psi_0 | \lambda K_0^{-1} |v\rangle\langle v| K_0^{-1} | \tilde\psi_0\rangle \det K_0\nonumber\\
	&= \lambda\langle v| K_0^{-2} | v\rangle \det K_0.
		\label{eq:instanton matrix algebra}
\end{align}
In the first line, we substitute statements (i) and (iii) into \eqref{eq:cofactor formula}; in the second, we note that $1-\lambda\langle v| K_0^{-1}| v\rangle = 0$ in our case, so only the second term of $K^{-1}$ gives any contribution \cite{footnote_instanton}. Finally, we use that $|\tilde\psi_0\rangle$ is the normalised zero mode, so by statement (ii), it must be $K_0^{-1}|v\rangle / \sqrt{\langle v|K_0^{-2}|v\rangle}$. Altogether, the measure of the instanton solutions relative to that of the instanton-free solution, $(\det K_0)^{-1/2}$, is
\begin{equation}
	\ud\tau \sqrt{\frac{\langle \psi | \psi\rangle}{2\pi\lambda \langle v| K_0^{-2} | v\rangle}} e^{-\Sa_\mathrm{cl}} .
	\label{eq:instanton measure final}
\end{equation}

In the $(k,\omega)$ basis, $K_0$ is positive definite and diagonal, $K_0(k,\omega) = \omega^2/\Da_0(k) + M$, $\lambda = 2\pi M/[\partial_\tau\phi_0(0,0)]$, and $v(k,\omega)=1$ (the Fourier transform of a $\delta$-function at the origin). It is easy to verify that $1-\lambda\langle v|K_0^{-1}|v\rangle = 0$, that is, $K$ indeed has a zero mode; furthermore, $K_0^{-1}|v\rangle \propto \partial_\tau \phi_0$, as expected. Substituting into \eqref{eq:instanton measure final} then gives that the measure of instanton solutions of a given sign is $\mu I$, where $I=e^{-\Sa_0}$ and
\begin{equation}
	\mu = \sqrt{M \pi \int (\ud^2 k) \omega(k)};
\end{equation}
at the RK point on the square lattice, $\mu = JS^{3/2} \sqrt{\pi/8}$.
%
%

\section{Details of the calculation for the honeycomb lattice}
\label{app: honeycomb details}

Following from Sec.~\ref{sec:Hexagonal-lattice}, 
we express the bosonic operators in the radial gauge
in the path integral formulation of the model. For $r \in A$: 
\begin{eqnarray}
b_{r,\eta}
&=&
\sqrt{\rho_\eta(r+e_\eta/2)}
  \exp\left[ i \Phi_\eta(r+e_\eta/2) \right]
\label{eq:radial_representation honeycomb}
\\
&\equiv&
\sqrt{\frac{S}{3}+\delta\rho_\eta(r+e_\eta/2)}
  \exp\left[ i \Phi_\eta(r+e_\eta/2) \right] 
\, . 
\nonumber 
\end{eqnarray}
For $r \in B$, the expressions are equivalent except for a change in sign, 
$+e_\eta/2 \to -e_\eta/2$. Notice that, by thinking of 
$\rho_\eta(r+e_\eta/2)$ and $\Phi_\eta(r+e_\eta/2)$ as functions defined on 
the midpoints of the bonds, there is no ambiguity nor redundancy in the 
notation. 

The constraints in Eq.~\eqref{eq:Constraint_S honeycomb} can then be written 
(choosing for concreteness $r \in A$) as 
\begin{eqnarray}
&& 
\sum_{\eta} \delta\rho_\eta(r+e_\eta/2) = 0 
\\ 
&&
\delta\rho_1(r+e_1/2) + 
\delta\rho_2(r+e_1-e_2/2) 
\nonumber \\* 
&&
\qquad \qquad \qquad \, 
+ \delta\rho_3(r+e_1-e_3/2) = 0 
\, . 
\end{eqnarray}
Taking the Fourier transform with respect to sublattice $A$, the constraints 
take on a more symmetric form:
\begin{align}
\sum_{\eta} e^{-i k e_\eta/2} \delta\rho_\eta(k) &= 0 ,
&
\sum_{\eta} e^{i k e_\eta/2} \delta\rho_\eta(k) &= 0 ,
\end{align}
where in the second line we divided out an overall factor $e^{-i k e_1}$. 
More conveniently, we can add and subtract them to obtain
\begin{align}
\sum_{\eta} c_\eta \, \delta\rho_\eta(k) &= 0 ,
&
\sum_{\eta} s_\eta \, \delta\rho_\eta(k) &= 0 ,
\end{align}
where again we used the shorthand notation $c_\eta = \cos(k e_\eta/2)$ 
and $s_\eta = \sin(k e_\eta/2)$. 
With three field variables $\delta\rho_\eta(k)$ and two constraints, we 
expect only one degree of freedom. 

As we did for the square lattice, it is convenient not to attempt to 
resolve the constraint directly but rather consider first the argument of 
the cosine term in the Hamiltonian, 
\begin{widetext}
\begin{eqnarray}
\tilde\phi(r) &=& 
\Phi_1(r+e_1/2) - \Phi_3(r+e_1-e_3/2) + \Phi_2(r+e_1-e_3+e_2/2) 
\nonumber \\* 
&-&\Phi_1(r+e_2-e_3+e_1/2) + \Phi_3(r+e_2-e_3/2) - \Phi_2(r+e_2/2),
\end{eqnarray}
and its Fourier transform 
\begin{eqnarray}
\tilde\phi(k) &=& 
\Phi_1(k)e^{-i k e_1/2} - \Phi_3(k)e^{-i k (e_1-e_3/2)} 
+ \Phi_2(k)e^{-i k (e_1-e_3+e_2/2)} 
\nonumber \\* 
&-&\Phi_1(k)e^{-i k (e_2-e_3+e_1/2)} + \Phi_3(k)e^{-i k (e_2-e_3/2)} 
- \Phi_2(k)e^{-i k e_2/2} 
\nonumber\\
&=& 
e^{i k e_3} 2 i \left[ 
  \Phi_1(k) (s_2 c_3 - s_3 c_2) 
  + \Phi_2(k) (s_3 c_1 - s_1 c_3) 
  + \Phi_3(k) (s_1 c_2 - s_2 c_1) 
\right] 
\, . 
\label{eq: tilde phi honeycomb}
\end{eqnarray}
\end{widetext}
The nicely symmetric expression in Eq.~\eqref{eq: tilde phi honeycomb} 
required a few lines of algebra and the property of the lattice vectors 
$e_1+e_2+e_3 = 0$. 
As we inferred above from the constraints on the $\delta\rho_\eta$ fields, 
there is only one real scalar degree of freedom, and as before it is 
convenient to do away with the phase factor [effectively, use plaquette centres as reference points for $\phi(r)$] and define 
\begin{eqnarray}
\phi(k) &=& e^{-i k e_3} \tilde\phi(k) 
\label{eq: phi(k) honeycomb}
\\*
&=& 
2 i \Big[ 
  \Phi_1(k) (s_2 c_3 - s_3 c_2) 
  + \Phi_2(k) (s_3 c_1 - s_1 c_3) 
\nonumber \\* 
&& \;\;
  + \Phi_3(k) (s_1 c_2 - s_2 c_1) 
\Big] 
\, . 
\nonumber 
\end{eqnarray}
[Notice the importance of the factor of $i$ in preserving the condition of 
Fourier transform of a real field, $\phi^*(k) = \phi(-k)$ due to the 
antisymmetric behaviour of the sine function.]

Finally, we look for a conjugate field $h(k)$ such that the Berry phase in 
the path integral can be written as $i h(k) \partial_\tau \phi(-k)$. 
Namely, we look for $h(k)$ that satisfies 
\begin{eqnarray}
h(k) \partial_\tau \phi(-k) 
= 
\sum_\eta \delta\rho_\eta(k) \partial_\tau \Phi_\eta(-k) 
\, . 
\end{eqnarray}
Substituting the expression for $\phi(k)$, 
Eq.~\eqref{eq: phi(k) honeycomb}, into the equation above, we obtain 
\begin{eqnarray}
\delta\rho_1(k) &=& -2i (s_2 c_3 - s_3 c_2) \, h(k) 
\nonumber \\* 
\delta\rho_2(k) &=& -2i (s_3 c_1 - s_1 c_3) \, h(k) 
\nonumber \\* 
\delta\rho_3(k) &=& -2i (s_1 c_2 - s_2 c_1) \, h(k) 
\, . 
\label{eq: honeycomb delta rho from height}
\end{eqnarray}
One can then straightforwardly verify that the introduction of the field 
$h(k)$ automatically resolves the constraints, 
\begin{eqnarray}
\sum_\eta c_\eta \, \delta\rho_\eta &\propto& 
\Big[ 
  c_1 (s_2 c_3 - s_3 c_2)
	+ c_2 (s_3 c_1 - s_1 c_3)
\nonumber \\*[-8pt]
	&+& c_3 (s_1 c_2 - s_2 c_1)
\Big] h(k) = 0 
\\[3pt]
\sum_\eta s_\eta \, \delta\rho_\eta &\propto& 
\Big[ 
  s_1 (s_2 c_3 - s_3 c_2)
	+ s_2 (s_3 c_1 - s_1 c_3)
\nonumber \\*[-8pt]
	&+& s_3 (s_1 c_2 - s_2 c_1)
\Big] h(k) = 0 
\, . 
\end{eqnarray}
We are thus in the position to write the full large-$S$ action of the system, 
including both the Berry phase and Hamiltonian contributions, in terms of 
the fields $h(k)$ and $\phi(k)$ only. 

To obtain the Gaussian field theory for the honeycomb lattice QDM, 
it is convenient to rewrite $\cos(\phi) = 1 - [1-\cos(\phi)]$ and 
notice that the term in square brackets contains only quadratic and higher-order contributions. 
Correspondingly, we can write
\begin{equation}
\sqrt{\rho\rho\rho\rho\rho\rho}
\cos(\phi)
\simeq 
\sqrt{\rho\rho\rho\rho\rho\rho}
-
\frac{S^3}{27} 
\left[1 - \cos(\phi)\right] ,
\label{eq: rho cos expansion honeycomb}
\end{equation}
and focus on expanding to second order the first term on the right hand side, 
as well as the $V\rho\rho\rho$ terms in the Hamiltonian in 
Eq.~\eqref{eq:RK_Hamiltonian honeycomb}. 
Linear terms in $\delta \rho_\eta(r+e_\eta/2)$ vanish upon summing over $r$ 
because of the dimer constraint. 
After quite some algebra, one arrives at the following 
contributions to quadratic order: 
\begin{eqnarray}
&&
\phantom{-{}}\frac{JS}{3} \Big\{ 
  \delta\rho_1(k) \delta\rho_1(-k) s_{23}^2
  +
  \delta\rho_2(k) \delta\rho_2(-k) s_{31}^2
  \nonumber \\*
  && \qquad \: 
  +
  \delta\rho_3(k) \delta\rho_3(-k) s_{12}^2
\Big\}
\label{eq: Hrho expansion honeycomb}
\\
&& 
-
\frac{JS}{3} \Big\{ 
  \big[
  \delta\rho_1(k)\delta\rho_2(-k)+\mathrm{c.c.} 
  \big] 
  c_{23} c_{31}
\nonumber \\*
&& \qquad \: 
  +
  \big[
  \delta\rho_2(k)\delta\rho_3(-k)+\mathrm{c.c.} 
  \big] 
  c_{31} c_{12}
\nonumber \\*
&& \qquad \: 
  +
  \big[
  \delta\rho_3(k)\delta\rho_1(-k)+\mathrm{c.c.} 
  \big] 
  c_{12} c_{23}
\Big\}
\nonumber \\
&& 
+
\frac{V S}{3} \Big\{
  \big[
  \delta\rho_1(k)\delta\rho_2(-k)+\mathrm{c.c.} 
  \big] 
  \big(c_{23} c_{31} + s_{23}s_{31}\big)
  \nonumber \\*
  && \qquad \: 
  +
  \big[
  \delta\rho_2(k)\delta\rho_3(-k)+\mathrm{c.c.} 
  \big] 
  \big(c_{31} c_{12} + s_{31}s_{12}\big)
  \nonumber \\*
  && \qquad \: 
  +
  \big[
  \delta\rho_3(k)\delta\rho_1(-k)+\mathrm{c.c.} 
  \big] 
  \big(c_{12} c_{23} + s_{12}s_{23}\big)
\Big\},
\nonumber
\end{eqnarray}
where we introduced 
\begin{eqnarray}
s_{\mu\nu} &=& \sin[k(e_\mu-e_\nu)/2] = s_\mu c_\nu - s_\nu c_\mu
\nonumber\\*
c_{\mu\nu} &=& \cos[k(e_\mu-e_\nu)/2] = c_\mu c_\nu + s_\mu s_\nu
\end{eqnarray}
for convenience.
Substituting the expressions \eqref{eq: honeycomb delta rho from height} of $\delta\rho_\eta(k)$ in terms of $h(k)$, 
and ignoring trivial constants, we obtain the action given in 
Eq.~\eqref{eq: action h,phi honeycomb} in Sec.~\ref{sec:Hexagonal-lattice}. 
%
%

\section{Details of the calculation for the diamond lattice}
\label{app: diamond details}

The interaction matrix $\Da_{\mu\nu}$ can be obtained in much the same way as for the honeycomb lattice, see Appendix.~\ref{app: honeycomb details}. For plaquettes in the $\mu=1,2,3$ sublattice, we obtain the contribution
\begin{widetext}
\begin{eqnarray}
&& 
\frac{JS}{4} \Big\{
  \delta\rho_1(k)\delta\rho_1(-k) s_{23}^2
 +\delta\rho_2(k)\delta\rho_2(-k) s_{31}^2
 +\delta\rho_3(k)\delta\rho_3(-k) s_{12}^2
\nonumber \\* 
&& \quad \: 
 -\big[
    \delta\rho_1(k)\delta\rho_2(-k)+\delta\rho_2(k)\delta\rho_1(-k) 
	\big]
	c_{23}c_{31}
 -\big[
    \delta\rho_2(k)\delta\rho_3(-k)+\delta\rho_3(k)\delta\rho_2(-k) 
	\big]
	c_{31} c_{12}
\nonumber \\* 
&& \quad \: 
 -\big[
    \delta\rho_3(k)\delta\rho_1(-k)+\delta\rho_1(k)\delta\rho_3(-k) 
	\big]
	c_{12}c_{23}
\Big\} 
\nonumber \\ 
&+& 
\frac{VS}{4} \Big\{
  \big[
    \delta\rho_1(k)\delta\rho_2(-k)+\delta\rho_2(k)\delta\rho_1(-k) 
  \big]
  \big(
	 c_{23} c_{31} + s_{23}s_{31}
  \big)
 +\big[
   \delta\rho_2(k)\delta\rho_3(-k)+\delta\rho_3(k)\delta\rho_2(-k) 
 \big]
 \big(
   c_{31} c_{12} + s_{31}s_{12}
 \big)
\nonumber \\ 
&& \quad \;\;
 +\big[
   \delta\rho_3(k)\delta\rho_1(-k)+\delta\rho_1(k)\delta\rho_3(-k) 
 \big]
 \big(
   c_{12} c_{23} + s_{12}s_{23}
 \big)
\Big\},
\end{eqnarray}
where we again introduced
$c_{\mu\nu}=\cos[k(e_\mu-e_\nu)/2]=c_\mu c_\nu+s_\mu s_\nu$ and 
$s_{\mu\nu}=\sin[k(e_\mu-e_\nu)/2]=s_\mu c_\nu-c_\mu s_\nu$. 
The other three sublattices of plaquettes give rise to equivalent contributions with
$123 \leftrightarrow 234 \leftrightarrow 341 \leftrightarrow 412$. 
Now, the matrix $\Da$ can be written explicitly as
\begin{equation}
\Da = \frac{JS}2\mathcal{J} + \frac{VS}2\mathcal{V}
\end{equation}
\begin{eqnarray}
	\mathcal{J} &=& \left(\begin{array}{cccc}
        s_{23}^2+s_{34}^2+s_{43}^2 & -(c_{13}c_{23}+c_{14}c_{24}) & -(c_{12}c_{32}+c_{14}c_{34}) & -(c_{12}c_{42}+c_{13}c_{43})  \\
        -(c_{23}c_{13}+c_{24}c_{14}) & s_{34}^2+s_{41}^2+s_{13}^2 & -(c_{21}c_{31}+c_{24}c_{34}) & -(c_{21}c_{41}+c_{23}c_{43}) \\
        -(c_{32}c_{12}+c_{34}c_{14}) & -(c_{31}c_{21}+c_{34}c_{24}) & s_{41}^2+s_{12}^2+s_{24}^2 & -(c_{31}c_{41}+c_{32}c_{42})\\
        -(c_{42}c_{12}+c_{43}c_{13}) & -(c_{41}c_{21}+c_{43}c_{23}) & -(c_{41}c_{31}+c_{42}c_{32}) & s_{12}^2+s_{23}^2+s_{31}^2
    \end{array}\right)\\[1ex]
    \mathcal{V} &=& \left( \begin{array}{*{3}{>{\centering}m{1.2in}|}>{\centering}m{1.2in}}
        0 &
        $c_{13}c_{23}-s_{13}s_{23}+c_{14}c_{24}-s_{14}s_{24}\phantom{{}+}$ & 
        $c_{12}c_{32}-s_{12}s_{32}+c_{14}c_{34}-s_{14}s_{34}\phantom{{}+}$ & 
        $c_{12}c_{42}-s_{12}s_{42}+c_{13}c_{43}-s_{13}s_{43}\phantom{{}+}$ 
        \cr\hline
        $c_{23}c_{13}-s_{23}s_{13}+c_{24}c_{14}-s_{24}s_{14}\phantom{{}+}$ & 
        0 &
        $c_{21}c_{31}-s_{21}s_{31}+c_{24}c_{34}-s_{24}s_{34}\phantom{{}+}$ & 
        $c_{21}c_{41}-s_{21}s_{41}+c_{23}c_{43}-s_{23}s_{43}\phantom{{}+}$ 
        \cr\hline
        $c_{32}c_{12}-s_{32}s_{12}+c_{34}c_{14}-s_{34}s_{14}\phantom{{}+}$ & 
        $c_{31}c_{21}-s_{31}s_{21}+c_{34}c_{24}-s_{34}s_{24}\phantom{{}+}$ & 
        0 &
        $c_{31}c_{41}-s_{31}s_{41}+c_{32}c_{42}-s_{32}s_{42}\phantom{{}+}$ 
        \cr\hline
        $c_{42}c_{12}-s_{42}s_{12}+c_{43}c_{13}-s_{43}s_{13}\phantom{{}+}$ & 
        $c_{41}c_{21}-s_{41}s_{21}+c_{43}c_{23}-s_{43}s_{23}\phantom{{}+}$ &
        $c_{41}c_{31}-s_{41}s_{31}+c_{42}c_{32}-s_{42}s_{32}\phantom{{}+}$ & 
        0
    \end{array} \right).
\end{eqnarray}
\end{widetext}
We can express the matrices $\mathcal{J}$ and $\mathcal{V}$ more concisely 
by introducing the matrix $\mathcal{C}$ with entries $c_{\mu\nu}$, the 
diagonal matrix $\mathcal{R}$ with entries 
$\mathcal{R}_{\mu\mu} = \sum_\nu \cos[k(e_\mu-e_\nu)]-2$, and defining 
$\Omega = \sum_{\mu<\nu} s_{\mu\nu}^2$: 
\begin{eqnarray}
	\mathcal{J} &=& \Omega\mathbb{I} + \mathcal{R} - \mathcal{C}^2+2\mathcal{C}\\
	\mathcal{V} &=& 2\mathcal{C}^2 - 6\mathcal{C} - \mathcal{R} ,
\end{eqnarray}
where $\mathbb{I}$ is the identity matrix. 

Ignoring instantons, the middle term of Eq.~\eqref{eq: action drho,Phi diamond} can be expanded about a given minimum of the cosine, giving 
\begin{eqnarray}
	&&\frac{JS^3}{64}\sum_{\alpha,r}  \phi_\alpha^2(r,\tau) = \frac{JS^3}{64}\sum_{\alpha,k} |\phi_\alpha(k,\tau)|^2 =
	\label{eq: diamond no-instantons phi2}\\
	&&
	\qquad 
	\frac{JS^3}{64}\sum_{k,\mu\nu\alpha} 
	  \mathcal{Z}_{\alpha\mu}(k)\mathcal{Z}_{\alpha\nu}(-k)
	  \Phi_\mu(k,\tau)\Phi_\nu(-k,\tau) 
		\, ,
  \nonumber
\end{eqnarray}
where 
$\mathcal{Z}_{\mu\nu} = 2i\varepsilon_{\mu\nu\lambda\kappa}s_\lambda c_\kappa$, 
see Eq.~\eqref{eq: phi_eta diamond}. 
Defining the matrix
\begin{equation}
\mathcal{M} =\frac{JS^3}{32} \mathcal{Z}^T(-k)\mathcal{Z}(k) = \frac{JS^3}{32}\mathcal{Z}^2,
\end{equation}
the quadratic action can now be written in the form \eqref{eq: diamond quadratic action}.

The dispersion $\omega^2(k)$ is given by the eigenvalues of 
the matrix $\mathcal{MD}$. It is important to note that 
$\mathcal{Z}_{\mu\nu}c_\nu = \mathcal{Z}_{\mu\nu}s_\nu=0$ by the definition 
of $\mathcal{Z}$. Consequently, $\mathcal{MC}=0$, which greatly simplifies 
the form of $\mathcal{MD}$. All in all, one has to diagonalise the matrix
\begin{equation}
	\frac{JS^4}{64} \big[\Omega J\mathcal{Z}^2 + (J-V)\mathcal{ZRZ}\big] 
	\, .
\end{equation}
It follows from the definition of $\mathcal{Z}$ that $c_\mu$ and $s_\mu$ 
(understood as 4-dimensional vectors) are eigenvectors of this matrix with 
zero eigenvalue. These correspond to the unphysical modes ruled out by the 
constraint~\eqref{eq: constraints delta rho diamond}. 
%
%

%
%


\begin{thebibliography}{10}
\bibitem{footnote_triplaq}
Curiously, in this case the condition that reduces the three 
cosine terms to two is the fact that one can decompose a hexagonal cell 
on the triangular lattice into 6 overlapping plaquettes; the product of the 
resulting $6$ plaquette-flipping operators amounts to the identity. Indeed, 
these $6$ plaquettes can be viewed as the $6$ faces of a cube projected onto a 
plane down one of its body diagonals.
\bibitem{footnote_instanton}
Note that, since $1-\lambda\langle v| K_0^{-1}| v\rangle = 0$ due to the presence of the zero mode, both terms in \eqref{eq:instanton matrix algebra 1} are ill-defined. However, both are finite for any other $\lambda$, so we can formally evaluate it elsewhere and take the limit of $\lambda$ going to its physical value. The divergence due to $1-\lambda\langle v| K_0^{-1}| v\rangle = 0$ cancels, so the correct limit is indeed given by \eqref{eq:instanton matrix algebra}.
\end{thebibliography}
\end{document}